\documentclass[aps,amsfonts,pra,onecolumn,superscriptaddress,showpacs,floatfix]{revtex4}

\usepackage{graphicx}
\usepackage{dcolumn}
\usepackage{bm}
\usepackage{natbib}
\begin{document}

\title{Finite one dimensional impenetrable Bose systems: Occupation numbers}

\author{P.J. Forrester}
\email[]{P.Forrester@ms.unimelb.edu.au}
\affiliation{Department of Mathematics and Statistics, University of Melbourne,
Victoria 3010, Australia}
\author{N.E. Frankel}
\email[]{n.frankel@physics.unimelb.edu.au}
\affiliation{School of Physics, University of Melbourne, Parkville,
Victoria 3010, Australia}
\author{T.M. Garoni}
\email[]{t.garoni@physics.unimelb.edu.au}
\affiliation{School of Physics, University of Melbourne, Parkville,
Victoria 3010, Australia}
\author{N.S. Witte}
\email[]{N.Witte@ms.unimelb.edu.au}
\affiliation{Department of Mathematics and Statistics, University of Melbourne,
Victoria 3010, Australia}
\affiliation{School of Physics, University of Melbourne, Parkville,
Victoria 3010, Australia}

\date{\today}
\newcommand{\norm}{\bar\lambda_0}
\newcommand{\sgn}{\text{\text{\text sgn}}}
\newcommand{\ds}{\displaystyle}
\newcommand{\ts}{\textstyle}
\newcommand{\be}{\begin{equation}}
\newcommand{\ee}{\end{equation}}
\newcommand{\ba}{\begin{eqnarray}}
\newcommand{\ea}{\end{eqnarray}}
\newcommand{\bi}{\bibitem}
\newcommand{\intl}{\int\displaylimits}
\newcommand{\suml}{\sum\displaylimits}
\newcommand{\prodl}{\prod\limits}
\newcommand{\cupl}{\cup\displaylimits}
\newcommand{\bl}{\left(}
\newcommand{\br}{\right)}
\newcommand{\wf}{\psi_N^C(x_1,x_2,...,x_N)}
\newcommand{\dm}{\varrho_{N}(t)}
\newcommand{\mom}{c_n(N)}
\newcommand{\dmf}{\varrho_N^F(t)}
\newcommand{\wdm}{\rho^H_N(x,y)}
\newcommand{\elem}{b_{j,k}(x,y)}
\newcommand{\w}{\sqrt{2N}}

\begin{abstract}
Bosons in the form of ultra cold alkali atoms can be confined
to a one dimensional (1d) domain by the use of harmonic traps.
This motivates the study of the ground state occupations $\lambda_i$
of effective single particle states $\phi_i$, in the theoretical 1d impenetrable Bose
gas. Both the system on a circle and the harmonically trapped system
are considered. The $\lambda_i$ and $\phi_i$ are the eigenvalues and
eigenfunctions respectively of the one body density matrix. We present
a detailed numerical and analytic study of this problem. Our main results are
the explicit scaled forms of the density matrices, from which it is
deduced that for fixed $i$ the occupations $\lambda_i$ are
asymptotically proportional to $\sqrt{N}$ in both the circular
and harmonically trapped cases.
\end{abstract}

\pacs{03.75.Fi, 05.30.Jp, 03.65.Ge}

\maketitle

\section{Introduction}
\label{introduction}
The Bose Gas, whose genesis we gratefully owe to Bose
and Einstein, is an icon of statistical physics. The
Bose-Einstein condensate (BEC) has found its  very special
place in nature as the fundamental mechanism underlying the properties
of a very special system, superfluid helium. In the past decade,
a new chapter has been opened with the experimental
realization of ultracold atomic Bose gases \cite{cornell,ketterle,pritchard}.

As such, new and intriguing possibilities for the Bose gas
have arisen. As these systems require harmonic traps, it
is possible to fashion them to constrain the bosons in low
dimensional configurations. In particular, Olshanii \cite{olshanii} has
pointed out that a one dimensional $(1d)$ configuration can
be obtained and maintained by making an elongated trap where one of the harmonic trap
frequencies, $\omega_z$, is much much less than the other frequency $\omega_\perp$.
The experimental realization of a $1d$  ultra cold atomic Bose
gas has now made this a physical reality\cite{dettmer,greiner,gorlitz}. 

This exciting prospect has now brought back into prominence
the noted quantum mechanical 
$N$-body problem, the impenetrable
1d Bose gas. This system was introduced by Girardeau \cite{girardeau} and
Lieb and Liniger \cite{lieb}. It was solved by Lenard in his two classic
works \cite{lenard,lenard66} and in the ensuing major works of Sutherland \cite{sutherlandjmp,sutherlandpra1,sutherlandpra2},
Vadiya and Tracy \cite{tracy,tracyphysrev,tracyerrata} and Jimbo et al \cite{jimbo}.

In \cite{lieb}, a $1d$ system of $N$ bosons  in box of length $L$ 
interacting via a  repulsive delta-function potential of strength, $g$,
was studied. It was shown that in the limit $\epsilon_F/g n \to 0$ the
impenetrable Bose gas was obtained, where $\epsilon_F$ is the Fermi energy and
$n=N/L$ is the number density. This limit corresponds to the condition $n a_1 \to 0$
where $g=\hbar^2/ma_1$ and $a_1$ is the $1d$ $s$-wave scattering length\cite{olshanii}.
To constrain the bosons to a $1d$ configuration the condition
$\epsilon_F/\hbar \omega_\perp \to 0$
must be obtained. This limit corresponds to the condition $n l_0 \to 0$
where $l_0 = \sqrt{(\hbar/ m\omega_\perp)}$, the characteristic range of the
harmonic trap. These and concomitant matters  are discussed in 
\cite{olshanii,dunjko,petrov}. Appendix \ref{energy appendix} gives the ground state energy and $\epsilon_F$ values
for the finite $N$ and large $N$ limit systems. 

The signature of the $1d$ impenetrable Bose system is the set of particle 
occupation numbers, which correspond  in the untrapped system to
the momentum distribution of the bosons but which  is quite different for the harmonically
trapped system. 

Section \ref{impenetrable bosons on a circle} presents our  comprehensive study of the
 occupation numbers for the finite $N$ and large $N$ limit  untrapped 
impenetrable boson systems. Section \ref{impenetrable bosons trapped in a
 harmonic well} with Appendix \ref{norm appendix}  presents our 
comprehensive study  of  the occupation numbers for the finite $N$ and
 large $N$ 
limit harmonically trapped boson systems. Section \ref{conclusion} presents concluding
remarks.

\section{Impenetrable bosons on a circle}
\label{impenetrable bosons on a circle}
\subsection{Properties of the density matrix}
\label{properties of the density matrix}
\subsubsection{The wave function}
\label{The wave function}
We are interested in the ground state of a one dimensional many body
Bose system which satisfies the free-particle Schr\"odinger equation, and also a condition of impenetrability which prevents two
bosons occupying the same point in space.
This system of impenetrable bosons is confined to lie
in a box of length $L$, and satisfy periodic boundary conditions,
i.e. 
\ba
\psi_N^C(x_1,...,x_k +L,...x_N)&=&\psi_N(x_1,...,x_k,...x_N),
\nonumber \\
k=1,2,...,N
\ea
which may be interpreted as confining the particles to move on a circle of
circumference length $L$. 

The impenetrability condition states that the wave function for our
$N$ body system must
vanish whenever two coordinates coincide, 
\be
\wf = 0 \quad{\text{\text{\text for}}}\quad x_j=x_k\quad (j\ne k).
\ee
For spin-less point particles however this condition is equivalent to
the Pauli exclusion principle. Hence for any fixed ordering of the
particles, e.g. 
\be
x_1<x_2<...<x_N,
\label{ordering}
\ee
there is no distinction between impenetrable bosons and  free
fermions\cite{girardeau}. Therefore for the ordering (\ref{ordering}), the ground state
of the system of impenetrable bosons can be constructed from a
Slater determinant of distinct single particle plane wave states, which has zero
total momentum and minimum total energy,

\ba
\wf
&=&
\cases{(L^NN!)^{-1/2}\det[e^{2\pi ikx_j/L}]_{j=1,...N\atop
k=-(N-1)/2,...,(N-1)/2} &N odd\cr
(L^NN!)^{-1/2}\det[e^{2\pi i(k+1/2)x_j/L}]_{j=1,...N\atop
k=-N/2,...,N/2} &N even\cr
}
\\
&=&
(L^NN!)^{-1/2}\prodl_{1\le j<k\le N}2i\sin[\pi(x_k-x_j)/L].
\label{realwavefunction} 
\ea

Ignoring irrelevant constant phase factors, we note that in the
region of configuration space which we are
considering, defined by condition (\ref{ordering}), the function (\ref{realwavefunction}) is
non-negative. This property of non-negativity distinguishes the ground
state in Bose systems. Imposing the condition that the wave function
for a system of bosons be symmetric under particle interchange,
$x_j\leftrightarrow x_k\quad(j\ne k)$, we
then deduce from (\ref{realwavefunction}) that for general ordering of
the particles, the properly normalized $N$ body wave function for our
system is 
\ba
\wf
&=&
(N!L^N)^{-1/2}
\prodl_{1\le j<k\le N}
2\vert\sin[\pi(x_k-x_j)/L]\vert
\\
&=&
(N!L^N)^{-1/2}
\prodl_{1\le j<k\le N}
\vert e^{2\pi i x_k/L} - e^{2\pi i x_j/L}\vert.
\label{wavefunction}
\ea
This result was first obtained by Girardeau \cite{girardeau}.

\subsubsection{Toeplitz determinant form for the density matrix}
\label{toeplitz determinant form for the density matrix}
The one body density matrix for a system of $N\ge 2$
particles is obtained by integrating out $(N-1)$ degrees of freedom from
the product 
$\psi_{N}(x_1,x_2,...,x_{N-1},x)\psi_{N}(x_1,x_2,...,x_{N-1},x')$. Since the periodic boundary conditions imply translational invariance
the one body density matrix will be a function of one variable
only. Specifically, for the wave function (\ref{wavefunction}) we have 
\ba
\rho_{N}^C(x-y)
&=&
N
\int_{0}^{L}dx_1...
\int_{0}^{L}dx_{N-1}
\,\psi_{N}(x_1,x_2,...,x_{N-1},x)
\psi_{N}(x_1,x_2,...,x_{N-1},y)
\label{dm definition}
\\
&=&
{1\over L}
{N\over N!}
\int_{0}^{2\pi}d\theta_1...
\int_{0}^{2\pi}d\theta_{N-1} 
\prodl_{l=1}^{N-1}
{1\over 2\pi}
\vert 1-e^{i\theta_l}\vert
\vert e^{2t i}-e^{i\theta_l}\vert
\prodl_{1\le j<k\le N-1}
\vert e^{i\theta_j} -e^{i\theta_k}\vert^2
\nonumber \\
&&
\label{intermediate toeplitz}
\ea
where $t = \pi(x-y)/L$, and to obtain (\ref{intermediate toeplitz}) we have made a
change of variables and utilized the periodicity of the integrand. The
normalization of (\ref{intermediate toeplitz}) is chosen so that 
%% \be
%% \rho_{N}(0)={N\over L}.
%% \label{normalisation}
%% \ee
It is convenient in to introduce a dimensionless
modification of $\rho_{N}^C(x)$ 
\be
\dm:=L\rho_{N}^C(x).
\label{circle notation}
\ee 

By utilizing the following general result it is straight forward to
express $\dm$ as the determinant of a Toeplitz
matrix \cite{szegotoeplitz}, a result first noted by Lenard \cite{lenard},
\be
{1\over N!}
\intl_{0}^{2\pi}d\theta_1
...
\intl_{0}^{2\pi}d\theta_N
\prodl_{l=1}^N f(\theta_l) 
\prodl_{1\le n<m\le N}
\vert e^{i\theta_j} -e^{i\theta_k}\vert^2
=
\det\left[\intl_{0}^{2\pi}d\theta f(\theta) e^{i\theta(j-k)}\right]_{j,k=1,2,...,N}.
\label{arousing}
\ee
The result (\ref{arousing}) can been obtained by identifying the 
double product as the squared modulus of the Vandermonde
determinant, which implies that the left hand side of (\ref{arousing}) is
\ba
{1\over N!}&&
\intl_{0}^{2\pi}d\theta_1
...
\intl_{0}^{2\pi}d\theta_N
\prodl_{l=1}^N f(\theta_l)
\sum_{P\in S_N}
\sum_{Q\in S_N}
\varepsilon(P)
\varepsilon(Q)
\prodl_{l=1}^{N}e^{i \theta_l[P(l)-Q(l)]}
\\
&=&
{1\over N!}
\sum_{P\in S_N}
\sum_{Q\in S_N}
\varepsilon(P)
\varepsilon(Q)
\prodl_{l=1}^{N}
\intl_{0}^{2\pi}d\theta\,f(\theta) 
e^{i \theta[P(l)-Q(l)]}
\label{penultimate heine identity step}
\\
&=&
\det\left[\intl_{0}^{2\pi}d\theta\,f(\theta)e^{i \theta(j-k)}\right]_{j,k=1,...N}
\label{final heine identity step}
\ea 
where $S_N$ is the group of all
permutations on $N$ objects, and $\varepsilon(P)$ is the signature of
the permutation $P$. In obtaining (\ref{final heine identity step}) from (\ref{penultimate heine identity step}) we have used the
general result that
\be
{1\over N!}
\sum_{P\in S_N}
\sum_{Q\in S_N}
\varepsilon(P)
\varepsilon(Q)
\prodl_{l=1}^{N}
a_{P(l),Q(l)}=\det\left[a_{j,k}\right]_{j,k=1,...N}.
\label{double sum equals det}
\ee

Applying (\ref{arousing}) to (\ref{intermediate toeplitz}) and
explicitly computing the one dimensional integrals we arrive at\cite{lenard}
\be
\dm=
\det\left[\gamma_{j-k}(t)
\right]_{j,k=1,...,N-1}
\label{toeplitz}
\ee
with $\gamma_n(t)$ given by
\ba
\gamma_n(t)
&=&
{1\over \pi}\int_{0}^{2\pi}
\vert\cos(u) -\cos\bl{t}\br\vert
e^{iun}\; du
\\
\gamma_n(t)
&=&
2\delta_{n,0}\cos(t) -
\delta_{n,1}-\delta_{n,-1}
-{4\cos(t)\sin(n\vert t\vert)\over \pi n}\nonumber
\\
&+&{2\over \pi}\left(
{\sin([n+1]\vert t\vert)\over (n+1)}+
{\sin([n-1]\vert t\vert)\over (n-1)}
\right).
\label{elements}
\ea
The succinct expressions (\ref{toeplitz}) and (\ref{elements}) are
well suited
for our numerical investigations of the momentum distribution, to be
discussed in Section \ref{numerical results}.

Expanding out the determinant (\ref{toeplitz})
we find explicitly
\ba
\varrho_2(t)
&=&
{2\over \pi}\left[(\pi -2t)\cos(t)+2\sin(t)\right],
\label{rhotwo}
\\
\varrho_3(t)
&=&
{1\over 2{\pi }^2}\left[\left(15 + 2{\left( \pi  - 2t \right) }^2 + 
4\left( 2 + \pi  - 2t \right)\left( \pi  - 2\left( 1 + t\right)\right)\cos (2t) + \cos (4t)\right)\right]
\nonumber \\
&+& 
{6\over \pi^2}\left( \pi  - 2t \right)\sin (2t),
\label{rhothree}
\\
\varrho_4(t)
&=&
{1\over 18{\pi }^3}\left[\left( 271 + 36{\left( \pi  - 2t  \right) }^2 \right)\left( \pi  - 2t  \right) 
\cos (t ) + 9\left( -33 + 4{\left( \pi  - 2t  \right) }^2 \right) 
\left( \pi  - 2t  \right) \cos (3t ) \right]
\nonumber \\
&+& 
{1\over 18{\pi }^3}\left[\left( \pi  - 2t  \right) \left( 25\cos (5t ) + \cos
(7t )
+ 
12\left( \pi  - 2t  \right) \left( 3\sin (t ) + 11\sin
(3t)\right)\right)\right]
\nonumber \\
&-& 
{1\over 18{\pi }^3}\left[
8\left(28\cos(2t) + 5\cos(4t)-105\right)
\sin^3(t)\right].
\label{rhofour}
\ea
For larger $N$, expanding out the determinant for $\dm$
quickly becomes unwieldy. We use the expressions
(\ref{rhotwo})-(\ref{rhofour}) 
in Section \ref{preliminaries} to find the corresponding exact 
expressions for the momentum distribution. 
\subsubsection{A differential equation for $\dm$}
\label{a differential equation for dm}
We present in this section a characterization of $\dm$ via a second order, second degree 
differential equation. This is a consequence of the identification\cite{cmp}
of $\varrho_{N+1}(t)$ with a certain well studied average over the eigenvalue
probability density function of the circular unitary ensemble of
$N\times N$ random matrices. It has recently been  demonstrated
\cite{peterandnick} that for a given $N$
such an average is expressible in terms of a specific $\tau$-function occurring
in the Hamiltonian formulation of the sixth Painlev\'e equation,
P$_{{\text{\text VI}}}$, and thus can be characterized by the
Jimbo-Miwa-Okamoto $\sigma$-form of this equation. 
Hence we obtain as a corollary an explicit differential equation
characterization for our $\dm$.  
Specifically, as a consequence of the detailed discussion in  \cite{cmp}
it can be seen that the function
\be
\phi_N(t)={(e^{2 i t}-1)\over 2i}{d\over dt}\log\varrho_N(t),
\label{phi}
\ee
satisfies the second order nonlinear differential equation
\ba
&&{\left( 1 - e^{-2i t} \right) }^2{\left( -2i \phi_N '(t)+ \phi_N ''(t) \right) }^2
=
16\left( 1 + \phi_N (t) + \frac{i }{2}\left( 1 - e^{-2i t} \right) \phi_N '(t) \right) 
\nonumber \\
&\times& 
\left( 2ie^{-2i t} \left( \phi_N (t) + \frac{i }{2}\phi_N '(t) \right) \phi_N '(t) + 
     (N-1)(N+1) \left( \phi_N (t) + \frac{i }{2}\left( 1 - e^{-2it} \right) \phi_N '(t) \right)  \right).
\nonumber \\
&&
\label{de}
\ea
The expression (\ref{phi}) can be inverted for $\dm$ to
yield 
\be
\varrho_N(t)
=
N
\exp \int_{0}^{t}ds\, {2 i\over e^{2si}-1}\phi_N(s).
\label{rhotophi}
\ee

If we substitute $u=e^{2it}$ into (\ref{de}), and define   
\be
\sigma_N(u)=\phi_N(t)=u(u-1){d\over
du}\log\varrho_N(t)\vert_{e^{2ti}=u},
\ee 
then the resulting equation is precisely of the Jimbo-Miwa-Okamoto $\sigma$-form
for  P$_{{\text{\text VI}}}$.

It is interesting to note \cite{cmp} that if in (\ref{phi}) we replace $\dm$ with
the one body density matrix for the free Fermi system,
$\varrho^F_N(t)$, 
then this
function is also a solution of the differential equation (\ref{de}), 
albeit with a different boundary condition.  
The boundary condition required to complete the characterization 
of $\dm$ or $\dmf$ in terms of (\ref{phi}) and (\ref{de}) is provided by knowing the first few terms of their small $t$
expansion. For $\dmf$ this is trivial since we simply have 
\be
\dmf
=
{\sin(N t)\over\sin(t)}
=N\left(
1-{(N^2-1)\over 6}t^2+
{(3N^4-10N^2+7)\over 360}t^4
+O(t^6)
\right).
\label{dmf}
\ee
To obtain the small $t$ expansion for $\dm$ we utilize the result, originally due to Lenard \cite{lenard66,cmp}, that 
$\rho_N^C(x)$ can be written in
terms of an expansion over integrals of the $n$-body density matrix of
the free Fermi system, which we
denote here by $\rho_N^F(x,x_2,x_3,...,x_n)\equiv
L^{-(n+1)}\varrho_N^F(t,t_2,t_3,...,t_n)$, with the obvious extension of
our notation $t_j=\pi x_j /L$. Recall that
$\varrho_N^F(t,t_2,t_3,...,t_{n+1})$ has
the following determinantal expression in terms of $\dmf$ 
\be
\varrho_N^F(t,t_2,t_3,...,t_{n+1})
=
\det
\pmatrix{
\dmf&[\varrho_N^F(t_j)]_{j=2,3,...,n+1}\cr
[\varrho_N^F(t-t_k)]_{k=2,3,...,n+1}&[\varrho_N^F(t_j - t_k)]_{j,k=2,3,...,n+1}\cr
}.
\label{ferminmatrix}
\ee
In the above, $j$ and $k$ refer to the row and column elements
respectively, so that for example $[\varrho_N^F(t_j)]_{j=2,3,...,n+1}$ is an $n$
dimensional row vector and $[\varrho_N^F(t_j - t_k)]_{j,k=2,3,...,n+1}$ is an
$n\times n$ matrix.

The one-body density matrix for the impenetrable Bose system then has
the form \cite{lenard66,cmp},
\be
\dm=
\sum_{n=0}^{\infty}{(-2)^n\over n!\pi^n}\int_{0}^{t}dt_2 ...\int_{0}^{t}dt_{n+1}\varrho_N^F(t,t_2,t_3,...,t_{n+1}).
\label{fermitobosefredholm}
\ee
We remark \cite{lenard66,cmp} that the right hand side of (\ref{fermitobosefredholm}) is exactly
proportional to the first Fredholm minor for the kernel
(\ref{dmf}). Using (\ref{dmf})-(\ref{fermitobosefredholm}), it is
straightforward to deduce that $\dm$ has the small $t$ behavior 
\be
{\dm\over \varrho_{N}(0)}=1 
- \frac{\left(  N^2-1 \right)}{6} t^2
+ \frac{N\left(  N^2 -1 \right)}{9\pi }t^3 +...
\label{boundary condition}
\ee
which then serves as the boundary condition characterizing $\dm$ in
terms of (\ref{phi})  and (\ref{de}).
\subsubsection{Asymptotic expansions of $\dm$, and the thermodynamic limit}
\label{asymptotic expansions of dm, and the thermodynamic limit}
One important
application of the differential equation  (\ref{de}), is
that it provides a far more expedient way of deriving the corrections
to the small $t$ expansion (\ref{boundary condition}) than the does the expression (\ref{fermitobosefredholm}).
By substituting a small $t$ power series ansatz for $\phi_N(t)$ into the
differential equation (\ref{de}), we obtain equations which define all but one of the 
coefficients. In particular the resulting equation for the coefficient of $t^3$
vanishes identically, and to fix this parameter we require the
boundary condition (\ref{boundary condition}) obtained from (\ref{fermitobosefredholm}). However the differential equation then provides
an extremely efficient way of obtaining a large number of higher order terms in
the small $t$ expansion. We found it straight forward to obtain the first twenty
terms of $\phi_N(t)$ in this manner, although it would require far too much space to
exhibit all these here. Using the relationship (\ref{phi}) we obtain
the corresponding small $t$ expansion of $\dm$, and include here terms up to and
including $t^9$,  
\ba
{\dm\over \varrho_{N}(0)}
&=&
1 
- \frac{\left(N^2 -1\right)}{6} t^2
+ \frac{N\left(N^2 -1\right)}{9\pi }t^3 
+ \frac{\left(3N^4 - 10N^2+ 7   \right)}{360} t^4
\nonumber \\
&-& 
  \frac{N\left( 11N^4 - 40N^2+ 29  \right)}{1350\pi }t^5 
- \frac{\left(3N^6 - 21N^4  + 49N^2 -31  \right)}{15120}t^6 
\nonumber \\ 
&+& 
  \frac{N\left(N^2 -1\right) \left(183N^4 - 1210N^2 +  2227\right) }{793800\pi } t^7
\nonumber \\
&+& \left(\frac{\left(N^2 -1\right)\left( 5N^6 - 55N^4 + 239N^2  -381\right)}{1814400} 
+ \frac{N^2\left( N^2  -4\right){\left(N^2 -1\right)}^2}{24300{\pi }^2}\right) t^8
\nonumber \\
&-& \frac{N\left(N^2 -1\right) \left(253N^6  - 3017N^4+ 13867N^2   -22863 \right)}{71442000\pi }t^9
+O(t^{10}).
\label{smallt}
\ea

Comparing this expansion with the expansion for $\dmf$ given in (\ref{dmf}), we see that while the
terms with an odd power of $t$ in (\ref{smallt}) arise purely from
the Bose nature of the system and are not present in the 
expansion of $\dmf$, the even terms in (\ref{smallt}) up to and
including $t^6$, are precisely just the corresponding free Fermi terms
from (\ref{dmf}). 
The coefficient of $t^8$ however is seen to be a composed of two pieces, the first is simply
the corresponding free Fermi term which is rational, where as the
second is irrational and dependent on the Bose nature of the system. Precisely the same behavior
is observed in the well known small $x$ expansion in the
thermodynamic limit \cite{jimbo}.

Our differential equation (\ref{de}) and  small $t$ expansion
(\ref{smallt}) for the finite system can be seen to reduce to the corresponding
results in the thermodynamic limit as follows. We define 
\be
\sigma_{\text V}(s)=\lim_{N\to\infty}\phi_N(s/N)=\lim_{N\to\infty}\sigma_N(e^{2is/N})
\ee
and denote the density matrix for the system in the thermodynamic
limit by $\rho(x)$. Taking $N\to\infty$ in (\ref{rhotophi}) and
(\ref{de}), and retaining only leading order terms, we respectively recover both the expression for 
$\rho(x)$ in terms of $\sigma_{\text V}(s)$, and the correct
$\sigma$-form of the fifth Painlev\'e equation characterizing $\sigma_{\text
V}(s)$, as discussed in the celebrated work of Jimbo et al\cite{jimbo}. Further, recalling that the thermodynamic limit
corresponds to $N\to\infty$ with the Fermi momentum $k_F=\pi N/L$ held
finite, we can replace $N t$ with $k_F x$ in (\ref{smallt}) and then noting that only the highest powers of $N$ in each
coefficient of (\ref{smallt}) will survive in this limit, we recover the corresponding
small $x$ asymptotic expansion for $\rho(x)$ due to Vaidya and Tracy\cite{tracy}.

In their tour de force \cite{tracy}, Vaidya and Tracy were also able to systematically construct the 
large $x$ asymptotic expansion for $\rho(x)$, the leading term of
which is 
\be
{\rho(x)\over \rho(0)}\sim\rho_{\infty} \vert k_F x\vert^{-1/2}.
\label{continuum large x}
\ee

The corresponding result for the finite system is
a good deal more subtle, as we deal with competing behavior of large $x$
and large $N$. 
Starting with the Toeplitz form (\ref{toeplitz}), a delicate analysis
due to Lenard\cite{lenardpacific}, rigorously justified by Widom\cite{widom}, gives the surprisingly simple result
\be
\varrho_N(t):=L\rho_N^C(x)\sim N \rho_{\infty}
\left\vert N \sin\left(k_F {x\over N}\right)\right\vert^{-1/2}
= \rho_{\infty}\sqrt{N}\vert\sin(t)\vert^{-1/2}
\label{large x expansion}
\ee
with
\be
\rho_{\infty}={G^4(3/2)\over\sqrt{2}}=\pi e^{1/2}2^{-1/3}A^{-6}\approx 0.92418
\ee
where $G(z)$ is the Barnes G function\cite{petersbook}, $A\approx 1.2824271$ is Glaisher's constant
and we have noted the identity \cite{petersbook,barnes}
\be
G(3/2)=\pi^{1/4}e^{1/8}2^{1/24}A^{-3/2}\approx 1.06922.
\ee
It is now known that (\ref{toeplitz}) is a special case of a class of
Toeplitz determinants with singular generating functions for which the
asymptotics can be computed (see for e.g. \cite{basor}).
We note too that in the thermodynamic limit the expression (\ref{large x
expansion}) reduces to (\ref{continuum large x}).
It is still an open question however, as to how to obtain the
correction terms to (\ref{large x expansion}) and thus generalize the
expansion of $\rho(x)$ of Vaidya and Tracy to the finite system. These correction terms
are important for investigating the large $N$ behavior of the momentum
distribution. 

\subsection{Momentum distribution}
\label{momentum distribution}
\subsubsection{Preliminaries}
\label{preliminaries}
Due to our choice of periodic boundary conditions the Fourier
coefficients of $\rho_N^C(x)$,
\be
\mom = \int_{0}^{L}dx\;
\rho_N^C(x)\exp\left({-2n\pi x i\over L}\right)
={1\over \pi}
\int_{0}^{\pi}dt\;
\dm\cos(2 nt),
\label{fouriercoefficients}
\ee
have the physical interpretation of being
the expectation of the number of particles in the single particle
state of momentum $2\pi n\hbar/L$.
In terms of these momentum state occupations, we have the Fourier
expansion for $\dm$ given by
\be
\dm
=
\sum_{n=-\infty}^{\infty}\mom\cos(2 nt).
\label{fourierexpansion}
\ee
Recalling our normalization convention, $\varrho_N(0)=N$, a direct consequence of (\ref{fourierexpansion}) is that
\ba
N
&=&\sum_{n=-\infty}^{\infty}\mom,
\label{momentum normalisation}\\
\varrho_N(\pi/2)
&=&\sum_{n=-\infty}^{\infty}(-1)^n\mom.
\label{obvious}
\ea 
The expression (\ref{momentum normalisation}) simply states
that the sum over the occupations of all momentum modes for our
system is equal to the total number of particles $N$.
It is of interest 
to compare the asymptotic occupation of the even modes
verses that of the odd modes, for large $N$. 
By subtracting (\ref{obvious}) from (\ref{momentum normalisation}),
and observing that (\ref{large x expansion}) implies
$\varrho_N(\pi/2)\sim \sqrt{N}$, we
obtain
\be
\sum_{n=-\infty}^{\infty}c_{2n+1}(N)={N\over 2}-{\varrho(\pi/2)\over 2}\sim {N\over 2}.
\label{odd}
\ee
Hence for large $N$ the total occupation of the odd modes
is half the total number of particles, and so in this limit the particles are evenly
distributed among both the odd and even modes.

It is straight
forward to obtain an exact expression for $c_n(2)$ for all $n$ by performing the integration (\ref{fouriercoefficients}) with
$\varrho_2(t)$ given by (\ref{rhotwo}), to obtain 
\be
c_n(2)
=
{16\over (4n^2-1)^2\pi^2}.
\label{c_n(2)}
\ee

We pause here to mention a mathematically interesting observation regarding the
structure  of the Fourier series
for $\varrho_2(t)$ at the special values 
$t=0$, and $t=\pi/2$. By definition we know $\varrho_N(0)=N$, and as 
discussed in \cite{cmp} the exact value of $\varrho_N(\pi/2)$ can be expressed in
closed form for all $N$. For $N=2$ the 
particular expressions (\ref{c_n(2)}) and (\ref{fourierexpansion}) yield  
\be
\varrho_2(0)={16\over \pi^2}E_1(2)=2\qquad {\text{\text{\text and}}}\qquad \varrho_2(\pi/
2)={16\over \pi^2}E_2(1)={16\over \pi^2}\arctan(1)={4\over \pi}
\ee
where $E_1(s)$ and $E_2(s)$ are the celebrated series studied by Euler\cite{ayoub,euler}
\be
E_1(s)=\sum_{k=0}^{\infty}{1\over (2k+1)^s}
=(1-2^{-s})\zeta(s)\qquad s>1, \qquad E_2(s)=\sum_{k=0}^{\infty}{(-1)^k\over (2k+1)^s},
\ee
which have the property 
that the sum of $E_1(s)$ for even $s\ge 2$,
and the sum of
$E_2(s)$ for odd $s\ge 1$ are both rational multiples of $\pi^{s}$.

It is possible to obtain exact closed form expressions of $\mom$ for
other small values of $N$, however the complexity of the corresponding
$\dm$ obviously prohibits us from taking this procedure too
far, and indeed already by $N=4$ the expression for $c_n(4)$ is quite
a complicated object. Explicitly,
\ba
c_n(3)
&=&
\cases{\displaystyle
{35\over 2\pi^2}\delta_{0,n}+{1\over 3}, &$\vert n\vert \le1$\cr
\displaystyle
{1\over 4 \pi^2}\delta_{\vert n\vert,2}
+\frac{6n^2 +2 }{n^2{\left(n^2 -1\right) }^2{\pi }^2}, & $\vert n\vert \ge1$\cr
}
\label{c_n(3)}
\\
c_n(4)
&=&
\frac{64\left( 80n^4  - 8n^2 +45 \right) }{{\left(16n^4 - 40n^2 + 9  \right) }^2{\pi }^2}
\nonumber \\
&+&  \frac{18874368\left(16n^2\left(26n^2\left(88n^2\left(21n^2-137\right)+15613\right)-179455\right)-55125\right)}
{{\left(16n^4-296n^2+1225\right)}^2{\left(16n^4-40n^2+9\right)}^4{\pi}^4}.
\label{c_n(4)}
\ea

\subsubsection{Large $N$ asymptotics of $\mom$}
\label{large $N$ asymptotics of $\mom$}
Utilizing the large $N$ asymptotic form for $\dm$
given by (\ref{large x expansion}), we can obtain the asymptotic behavior of $\mom$ for
 $N\gg n$,
\be
\mom\sim  {\rho_{\infty}\over\pi} \sqrt{N}
\int_{0}^{\pi}dt \cos(2nt)\sin^{-1/2}(t).
\ee
This integral is expressible in terms of
the beta function \cite{prud}, $B(x,y)$, and so
\ba
\mom
&\sim& 
 {\rho_{\infty}\over\pi}
{2\sqrt{2}\pi\cos(n\pi)\over B(n+3/4,-n+3/4)}\sqrt{N}
\\
\mom
&\sim&
{\rho_{\infty}\over \sqrt{\pi}} {\Gamma(n+1/4)\over\Gamma(n+3/4)} \sqrt{N}.
\label{gamma prod}
\ea
Thus the leading order large $N$ behavior of $c_0(N)$ is 
\be
c_0(N)
\sim
{\sqrt{2\pi}\rho_{\infty}\over\Gamma^2(3/4)} \sqrt{N}
\approx 1.54269 {\sqrt{N}}\qquad N\to\infty .
\label{approxc_0}
\ee
A consequence of our lack of correction terms to the asymptotic result (\ref{large
x expansion}) for $\dm$, is that only the leading order large $N$ behavior of
$\mom$ is available to us via this argument. In Section
\ref{numerical results} we use numerical results to conjecture the
form of the correction terms to (\ref{gamma prod}).
 
Manipulating (\ref{gamma prod}) we find that  the ratio of gamma
functions
\footnote{We wish to observe that this ratio of gamma functions has been studied in
the celebrated work on continued fractions of Stieltjes\cite{stieltjes}
and Ramanujan\cite{ramanujan}.}
reduces to a simple rational product and thus we obtain the following very tidy  expression for the large $N$
behavior of $\mom$ in
terms  of that of $c_0(N)$ as follows
\be
{\mom\over c_0(N)}\sim  \prod_{l=1}^{n}{(4l-3)\over (4l-1)}.
\label{cn on c0 exact}
\ee 

If $n$ also is asymptotically large, so that $N\gg n\gg 1$, we can
simplify things further still by
applying 
Stirling's formula to (\ref{gamma prod}) which yields 
\ba
\mom
&\sim& 
{\rho_{\infty}\over\sqrt{\pi}}{\sqrt{N}\over
\sqrt{n}}\left[1-{1\over 64n^2}+O(n^{-3})\right],
\label{large N and large n}
\\
\Longrightarrow {\mom\over c_0(N)}
&\sim&
{\Gamma^2(3/4)\over \sqrt{2}\pi}{1\over \sqrt{n}}\left[1-{1\over 64n^2}+O(n^{-3})\right].
\label{cn on c0 stirling}
\ea
It is important to observe  that there are no corrections linear in $n^{-1}$
in the expressions (\ref{large N and large n}) and (\ref{cn on c0
stirling}), which is a consequence of the fact that the first
correction terms in the Stirling expansions of $\Gamma(n+1/4)$ and
$\Gamma(n+3/4)$ are identical. The practical consequence of this is
that (\ref{cn on c0 stirling}) provides  an extremely good approximation of (\ref{cn on c0 exact})
even for small $n$, in particular even for $n=1$. To demonstrate this remarkable
agreement we list in Table \ref{table2} the values of
${\mom/c_0(N)}$ evaluated via both (\ref{cn on c0 exact}) and (\ref{cn on c0
stirling}), for $n=1,2,3,5$. 

\begin{table*}
\begin{tabular}{|c|c|c|} \hline
$n$ &  $\mom/c_0(N)$ evaluated using (\ref{cn on c0 exact}).&$\mom/c_0(N)$ evaluated using (\ref{cn on c0 stirling}). \\
\hline\hline
$1$ &  $1/3=0.333333...$                 & $0.332708...$ \\
\hline
$2$ &  $5/21=0.238095...$                & $0.238061...$ \\
\hline
$3$ &  $15/77=0.194805...$               & $0.194799...$  \\
\hline
$5$ &  $221/1463=0.151059...$               & $0.151059...$  \\
\hline
\end{tabular}
\caption{Comparison of the values for  $\mom/c_0(N)$ obtained via
(\ref{cn on c0 exact}) versus (\ref{cn on c0 stirling}).}
\label{table2}
\end{table*}

Finally, we compare our results for the finite
system when $N\gg n>0$ with the continuum momentum distribution
constructed using the  large $x$ expansion (\ref{continuum large x})
for $\rho(x)$ of Vaidya and
Tracy\cite{tracy}. Denoting the
momentum distribution in the thermodynamic limit by $c(k)$, we find
the following small $\vert k\vert$ behaviour
\be
c(k)=\int_{-\infty}^{\infty}e^{-ikx}\rho(x)\, dx
\sim \rho_{\infty}\int_{-\infty}^{\infty}e^{-ikx}\vert k_F x\vert^{-1/2}\, dx
=\sqrt{{2\over\pi}}\rho_{\infty}
\left\vert {k\over k_F} \right\vert^{-1/2}.
\label{jimbo momentum}
\ee
Substituting $k=2\pi n/L$, consistent with our periodic boundary
conditions, into (\ref{jimbo momentum}) results in the leading term of
(\ref{large N and large n}). This is
consistent since even with $n \gg 1$, if $N\gg n$ this implies that
$k\ll k_F$. 

\subsubsection{Numerical calculation of $c_0(N)$, $c_1(N)$ and $c_2(N)$}
\label{numerical results}

\begin{figure}
\includegraphics{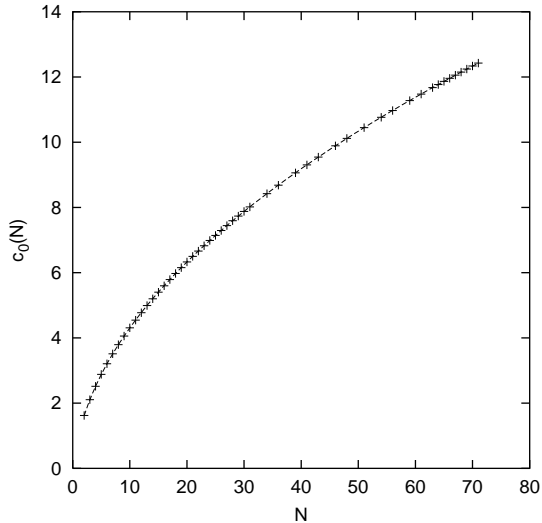}
\caption{The zeroth mode of the momentum distribution $c_0(N)$ vs $N$.} 
\label{zerothmode}   
\end{figure}

\begin{figure}
\includegraphics{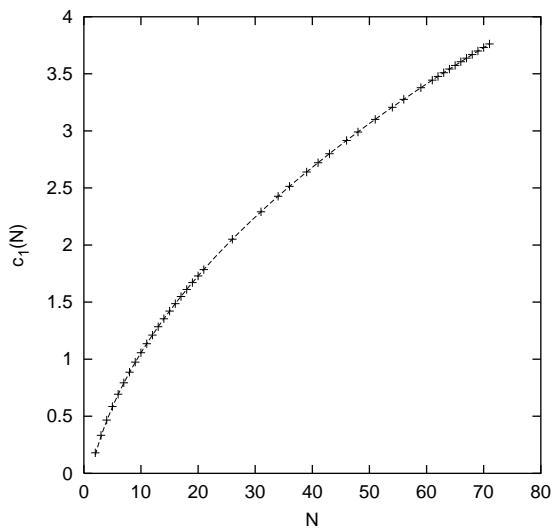}
\caption{The first mode of the momentum distribution $c_1(N)$ vs $N$.} 
\label{firstmode}   
\end{figure}

\begin{figure}
\includegraphics{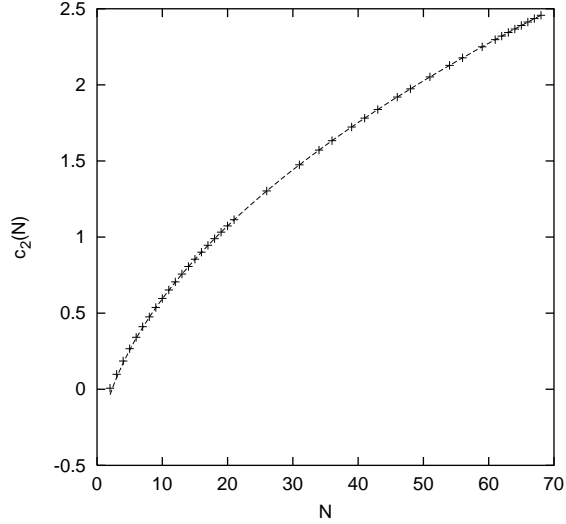}
\caption{The second mode of the momentum distribution $c_2(N)$ vs $N$.} 
\label{secondmode}   
\end{figure}

In order to gain further insight into the occupations of the low lying
momentum  modes, we have performed the Fourier integral (\ref{fouriercoefficients}) over
the Toeplitz determinant (\ref{toeplitz}) numerically, for various
values of $N$, for the cases $c_0(N)$, $c_1(N)$, and $c_2(N)$. The
results are shown in Figs. \ref{zerothmode}, \ref{firstmode} and
\ref{secondmode}, where the dots
represent the result of performing the numerical integration and
the line represents a fit of this data to the ansatz 
\be
\mom \sim a_1\sqrt{N}+a_2 +a_3N^{-1/2}+a_4N^{-1}+...
\label{ansatz}
\ee
We also list in Table \ref{table1} the exact values of $c_0(N)$ for
$N=2,3,4,5,6,7$.
\begin{table*}
\begin{tabular}{|c|c|c|} \hline
$N$ &  Exact $c_0(N)$                           & Numerical $c_0(N)$ \\
\hline\hline
$2$ &  $\frac{16}{{\pi }^2}$                   & $1.62114...$ \\
\hline
$3$ &  $\frac{1}{3} + \frac{35}{2\,{\pi }^2}$                & $2.10645...$ \\
\hline
$4$ &  $\frac{-2097152}{19845\,{\pi }^4} + \frac{320}{9\,{\pi }^2}$               & $2.51766...$  \\
\hline
$5$ &  $\frac{1}{5} + \frac{7436429}{129600\,{\pi }^4} + \frac{4459}{216\,{\pi }^2}$           & $2.88069...$  \\
\hline
$6$ &  $\frac{193507848058308060419981312}{12748157814913474078125\,{\pi }^6}- 
        \frac{38494793629696}{21739843125\,{\pi }^4} + 
        \frac{4144}{75\,{\pi }^2}$             &$3.20923...$  \\
\hline
$7$ &  $\frac{1}{7} + 
        \frac{85760621135804297813}{40663643328000000\,{\pi }^6} - 
        \frac{46891706849}{317520000\,{\pi }^4} + 
        \frac{79679}{3000\,{\pi }^2}$          &$3.51155...$   \\       
\hline
\end{tabular}
\caption{Values of $c_0(N)$ for $N=2,3,4,5,6,7$}
\label{table1}
\end{table*}

As discussed in Section
\ref{large $N$ asymptotics of $\mom$}, analytic arguments lead to
$\sqrt{N}$ leading order behavior for the large $N$
asymptotics of $\mom$. The coefficients of $\sqrt{N}$ for
$c_0(N)$, $c_1(N)$, and $c_2(N)$ given in our numerical
fits below match very well with those  derived by such
arguments.
\ba
c_0(N)\approx 1.54273\,{\sqrt{N}} - 0.5725 + \frac{0.003677}{{\sqrt{N}}} 
\label{c_0numerical}
\\
c_1(N)\approx 0.514345\,{\sqrt{N}}  -0.5739 +\frac{0.01128}{{\sqrt{N}}}
\label{c_1numerical}
\\
c_2(N)\approx 0.367622\,{\sqrt{N}}-0.5775  + \frac{0.02948}{{\sqrt{N}}}.
\label{c_2numerical}
\ea

The fits (\ref{c_0numerical})-(\ref{c_2numerical}) were
constructed using only the data for $N> 30$ so as to obtain 
asymptotic information for large $N$, however as can be seen from the
plots they match up very well with all the data. Indeed, the
$\chi^2$ value for each fit is of the order $10^{-10}$. We note that
the coefficient of the $N^{-1/2}$ term in each of
(\ref{c_0numerical})-(\ref{c_2numerical}) is considerably smaller
than the coefficient of the leading term or the constant. Also we
remark that the constant term is very similar in all three fits, and
that a very similar constant again appears in the fits of the occupation
numbers for the system of harmonically trapped impenetrable bosons to
be discussed in Section \ref{numerical results}. This 
suggests that the first correction term to the the occupation of the
low lying states as a function of $N$ may well be a universal constant common
to both of these systems.
\subsubsection{Large $n$ asymptotics of $\mom$}
\label{large $n$ asymptotics of $\mom$}
The occupation numbers, $\mom$, depend on the two parameters $n$ and
$N$. In Section \ref{large $N$ asymptotics of $\mom$} we found the
asymptotic behaviour of $\mom$ for $N\gg n$. 
In this section, by contrast, we seek an expression for the large $n$ behaviour of
$\mom$, with $n\gg N$. Observing the large $n$ behaviour of the
exact results for $c_n(2)$, $c_n(3)$ and $c_n(4)$, given respectively by (\ref{c_n(2)}), (\ref{c_n(3)})
and (\ref{c_n(4)}), we are led to develop the expansion 
\be
\mom = {a_1\over n^4}+{a_2\over n^6}+...
\label{large n ansatz}
\ee
To obtain the coefficients $a_1, a_2,..$ we construct a small $t$ expansion of $\dm$
by applying the Mellin transform technique to its Fourier series
representation (\ref{fourierexpansion}) with Fourier coefficients
given by (\ref{large n ansatz}), and then compare this result
with (\ref{smallt}).
Beginning with (\ref{fourierexpansion}) we insert the Mellin
integral representation of $\cos(2nt)$ which yields
\be
\dm 
=
c_0(N)
+{1\over 2\pi i}\int_{c-i\infty}^{c+i\infty}ds\,
2^{1-s} \Gamma(s) \cos(\pi s/2) g(s) t^{-s},\qquad 0<c<1
\label{mellinrho}
\ee
where we have defined 
\be
g(s)=\sum_{n=1}^{\infty}{\mom\over n^s}=\sum_{j=1}^{\infty}a_j\zeta(s+2j+2)
\label{g}
\ee
and in the last step we have swapped orders of summation and
identified the series representation of the Riemann zeta function,
$\zeta(s)$, which is finite everywhere on our contour of integration
in (\ref{mellinrho}).

Closing the contour of (\ref{mellinrho}) to the left results in a
small $t$ expansion. The poles of the integrand  in the left half
plane are all simple, and lie at $s=0,-2,-3,-4,-5,...$. We note that the term arising
from the residue at $s=0$ combines with $c_0(N)$ to produce
$\varrho_N(0)$. 
Since $\Gamma(s)\cos(\pi s/2)$ is analytic when $s$ is odd, the
poles at  $s=-(2k+1)$, $k\ge 1$, arise purely from the $a_k$ term
in the zeta function expansion of $g(s)$. Hence the corresponding $t^{2k+1}$
term in the small $t$ expansion of $\dm$ will contain $a_{k}$, so that
we can obtain the values of all the $a_k$ simply by considering the odd
terms. We note that as discussed
in Section \ref{asymptotic expansions of dm, and the thermodynamic
limit}, it is precisely the odd terms in the small $t$ expansion of
$\dm$ which arise purely from the Bose nature of the system and do not
appear in the corresponding free Fermi expansion. The
poles at even values of $s$ arise
from $\Gamma(s)$, since $g(s)$ is analytic then, and so the residues at these values will contain
the entire $g(s)$ series, (\ref{g}). Calculating the required residues then, we obtain the following expansion
\ba
{\dm\over \varrho_N(0)}
&\sim&
1-{4g(-2)\over N}t^2 +{4\pi\over 3N}a_1 t^3 +{4g(-4)\over 3N}t^4
-{4\pi\over 15N}a_2 t^5-{8g(-6)\over 45N}t^6
\nonumber \\
&+&
{8\pi\over 315(N+1)}a_3 t^7 + \frac{4g(-8)}{315N}t^8 -
\frac{4\pi}{2835N}{a_4}t^9.
\label{mellinexpansion}
\ea
 Comparing (\ref{mellinexpansion}) with
(\ref{smallt}) we deduce
\ba
\mom
&\sim&
  \frac{N^2\left( N^2 -1 \right) }{12{\pi }^2}{1\over n^4} 
+ \frac{N^2\left( N^2 -1 \right) \left( -29 + 11N^2 \right)
 }{360{\pi}^2} {1\over n^6}
\nonumber \\
&+& 
\frac{N^2\left( N^2 -1 \right) \left( 2227 - 1210N^2 + 183N^4\right)
 }{20160{\pi }^2}{1\over n^8} 
\nonumber \\
&+&
 \frac{N^2\left( N^2 -1 \right) \left( -22863 + 13867N^2 - 3017N^4 +
 253N^6 \right) }{100800{\pi }^2}{1\over n^{10}}.
\label{large n}
\ea

Going out to higher orders is a straight forward matter, but the resulting
expressions obtained for $a_k$ become increasingly cumbersome.

Feeding the explicit
values of $a_k$ from (\ref{large n}) into $g(s)$ and comparing
the even terms in (\ref{mellinexpansion}) with the
corresponding even terms in (\ref{smallt}) we  
obtain a consistency check on the validity of our ansatz. 

We note that for the values $N=2,3,4$, the large $n$ expansion
(\ref{large n}) recovers the large $n$ expansions one obtains from the
exact exact results (\ref{c_n(2)}), (\ref{c_n(3)}) and  (\ref{c_n(4)}).     
Further, we can also recover from (\ref{large n}) the large $k$
behaviour of the continuum momentum distribution, $c(k)$, just as we
recovered the small $k$ behaviour of $c(k)$ from (\ref{large N and large
n}) in Section \ref{large $N$ asymptotics of $\mom$}.  
If we apply the following asymptotic result\cite{wong}
\ba
\rho(x)
&\sim& 
\rho(0)\sum_{s=0}^{\infty} b_s (k_Fx)^s,
\qquad x\to 0
\\
\Longrightarrow 
\int_{0}^{\infty}\rho(x)\cos(kx)\, dx
&\sim&
{1\over \pi}\sum_{s=0}^{\infty}(-1)^{s+1} (2s+1)!  b_{2s+1} \left({k_F\over k}\right)^{2(s+1)},
\qquad k \to\infty
\ea
to the small $x$ asymptotic expansion for $\rho(x)$ of Vaidya and Tracy\cite{tracy}
\ba
{\rho(x)\over \rho(0)}
&=&
1-{(k_Fx)^2\over 6}+{\vert k_Fx\vert^3\over 9\pi}
+{(k_Fx)^4\over 120}-{11\vert k_Fx\vert^5\over 1350\pi}-{(k_Fx)^6\over 5040}+{122\vert k_Fx\vert^7\over 105\pi7!}
\nonumber \\
&+&
\left({1\over 24\,300\pi^2} +{1\over 9!}\right)(k_Fx)^8 -{253\over
98\,000\pi 27^2}\vert k_F x\vert^{9} +O((k_Fx)^{10})
\ea
we deduce that
\be
c(k)\sim
{4\over 3\pi^2}\left({k_F\over k}\right)^4+{88\over
45\pi^2}\left({k_F\over k}\right)^6
+{244\over 105\pi^2}\left({k_F\over k}\right)^8
+\frac{4048}{1575\,{\pi }^2}\left({k_F\over k}\right)^{10}.
\label{large k tdl}
\ee
Substituting $k=2\pi n/L$, which implies $N/n=2k_F/k$, into
(\ref{large n}) and taking $N\to\infty$ retaining only leading order
terms then recovers (\ref{large k tdl}). 

We can thus summarise the asymptotic behaviour  of the occupation of
the $n$th single particle momentum state for a system of impenetrable
bosons on a circle as follows: for large $N$ and fixed $n$,  $\mom$
diverges as $\sqrt{N}$ , whilst it tends to zero like $n^{-4}$ when $n\gg N$. 
As will become evident in Section \ref{impenetrable bosons trapped in
  a harmonic well}, this summary also describes the behaviour of a
system of harmonically trapped impenetrable bosons on a line.

\section{Impenetrable bosons trapped in a harmonic well}
\label{impenetrable bosons trapped in a harmonic well}
\subsection{The analytic structure of the density matrix}
\subsubsection{The wave function}
The wave function for a system of $N$ impenetrable bosons on a line confined by a harmonic
potential is defined by  the Hamiltonian (in reduced units)
\be
-\sum_{j=1}^{N} {\partial^2\over \partial x_j^2}+\sum_{j=1}^{N}x_j^2
\label{well hamiltonian}
\ee
and the impenetrability condition discussed in Section (\ref{The wave
  function}). The well known  single
particle eigenstates of (\ref{well hamiltonian}) have the form
\be
{e^{-x^2/2}H_j(x)\over \sqrt{\pi}2^j j!}\qquad j=0,1,2,...
\label{single pcle well states}
\ee
where $H_j(x)$ is the $j$th Hermite polynomial. 
The ground state wave function
for the harmonically trapped non interacting Fermi system is
obtained by forming the Slater determinant of the functions
(\ref{single pcle well states}) with $j=0,1,...,N-1$. 
Arguing then as in Section (\ref{The wave function}) we obtain
the ground state wave function for the impenetrable bosons by simply
taking the modulus of this Slater determinant. 

The Vandermonde determinant formula states that
\be
\det[p_{j-1}(x_k)]_{j,k=1,...N}=\prod_{1\le j,k\le N}(x_k-x_j)
\label{vandermonde}
\ee
for any set $\{p_j(x)\}$ with $p_j(x)$ a monic polynomial of degree
$j$, i.e. a polynomial of degree $j$ for which the coefficient of
$x^j$ is $1$. In particular,
(\ref{vandermonde}) is true for  $\{p_j(x)\}=\{2^{-j}H_j(x)\}$ and $\{p_j(x)\}=\{x^j\}$. 
Applying (\ref{vandermonde}) to the Slater determinant over
(\ref{single pcle well states}) we see that the ground state wave function for $N$
harmonically trapped impenetrable bosons can be expressed as  
\be
\psi^{H}_{0}(x_1,x_2,...,x_N)
=
{1\over C_N^H}\prod_{k=1}^{N}e^{-x_k^2/2}\prod_{1\le j<k \le N}\vert
x_j -x_k\vert,
\label{well wave function}
\ee
where
\be
(C_N^H)^2
=
N!\prod_{m=0}^{N-1}2^{-m}\sqrt{\pi}m!.
\nonumber
\ee
\subsubsection{The one body density matrix}
The one body density matrix is defined as in (\ref{dm definition})
with the domain of integration now ${\mathbb R}$,
\be
\wdm=
N\int_{-\infty}^{\infty}dx_1...\int_{-\infty}^{\infty}dx_{N-1}
\psi_N^H(x_1,...x_{N-1},x)\psi_N^H(x_1,...x_{N-1},y).
\label{well multidimensional integral}
\ee
and is now genuinely a function of two variables.
We are again able to find a closed form
expression for $\wdm$ in terms of a determinant, analogous to the
result on the circle, this time the determinant being of Hankel type
rather than Toeplitz type. To make this identification it is
advantageous to note the following general result, in analogy to
(\ref{arousing})
\be
{1\over
  N!}\prod_{l=1}^{N}\int_{-\infty}^{\infty}dx_l\;g(x_l)\left(\det[f_{j-1}(x_k)]_{j,k=1,...,N}\right)^2
=
\det\left[\int_{-\infty}^{\infty}dt\;g(t) f_{j-1}(t) f_{k-1}(t)\right]_{j,k=1,...,N},
\label{well arousing}
\ee
which can be obtained 
by simply expanding the determinants and recalling (\ref{double sum
  equals det}) so  that left hand side becomes
\ba
&&
{1\over N!}
\sum_{P\in S_N}
\sum_{Q\in S_N}
\varepsilon(P)
\varepsilon(Q)
\prod_{l=1}^{N}
\int_{-\infty}^{\infty}dx_l\;g(x_l)f_{Q(l)-1}(x_l)f_{P(l)-1}(x_l)
\label{penultimate well arousing}
\\
&=&
\det\left[\int_{-\infty}^{\infty}dt\;g(t) f_{j-1}(t) f_{k-1}(t)\right]_{j,k=1,...N}.
\label{final well arousing}
\ea 

To obtain a closed
form for $\wdm$ we proceed as follows. Consider
$\psi_N^H(x_1,...,x_{N-1},x)$ from (\ref{well wave function}) and factor out the pieces that depend
upon $x$, then apply (\ref{vandermonde}) to obtain 
\be
\wdm
=
{N\over
  C_N^2}e^{-x^2/2-y^2/2}\prod_{l=1}^{N-1}\int_{-\infty}^{\infty}dx_l\;e^{-x_l^2}\vert
  x-x_l\vert\vert y -x_l\vert
(\det\left[x_k^{j-1}\right]_{j,k,1,...N-1})^2
\ee
and hence \cite{cmp}
\be
\wdm=
 {2^{N-1}\over \sqrt{\pi}\Gamma(N)}e^{-x^2/2-y^2/2}
\det\left[{2^{(j+k)/2}\over 2\sqrt{\pi}\sqrt{\Gamma(j)\Gamma(k)}}\elem\right]_{j,k=1,...N-1}
\label{well dm}
\ee
where in the last step use has been made of (\ref{well arousing}). 
The elements of the determinant have the following explicit form \cite{cmp}
\ba
\elem
&=&
\int_{-\infty}^{\infty}
dt\; e^{-t^2}\vert x-t\vert\vert y-t\vert  t^{j+k-2}\qquad 1\le j,k\le N-1
\label{well element integral}
\\
\elem
&=&
\int_{-\infty}^{\infty}
dt\; e^{-t^2}(x-t) (y-t)  t^{j+k-2}
\nonumber \\
&-&
2\,\sgn(y-x)\int_{x}^{y}dt\;e^{-t^2}(x-t) (y-t)  t^{j+k-2}
\label{well fermi bose integral comparison}\\
%% xy\theta_{j+k-1} -(x+y)\theta_{j+k} +\theta_{j+k+1}
\elem
&=&
f_{j,k}(x,y)
\nonumber \\
&-&
2\,\sgn(y-x)
[xy\mu_{j+k-2}(x,y)-(x+y)\mu_{j+k-1}(x,y)+\mu_{j+k}(x,y)]
\label{well fermi bose comparison} 
\ea
where for clarity we have introduced
\ba
f_{j,k}(x,y)
&:=&
\int_{-\infty}^{\infty}
dt\; e^{-t^2}(x-t) (y-t)  t^{j+k-2}
\\
&=&
\cases{\Gamma\left({j+k-1\over 2}\right)xy
  +\Gamma\left({j+k+1\over 2}\right)& j+k even\cr
-\Gamma\left({j+k\over 2}\right)(x+y)& j+k odd}
%% \theta_m
%% &=&
%% {[1+(-)^{m+1}]\over 2}\Gamma\left({m\over 2}\right)
\\
\mu_m(x,y)
&:=&
\int_x^y dt\; e^{-t^2}t^m 
\\
&=&
{(\sgn(y))^{m+1}\over 2}\gamma\left({m+1\over 2},y^2\right)
-{(\sgn(x))^{m+1}\over 2}\gamma\left({m+1\over 2},x^2\right)
\nonumber \\
&=&
{y^{m+1}e^{-y^2}\over (m+1)} {_1F_1}\left(1,{m+3\over 2},y^2\right)
-
{x^{m+1}e^{-x^2}\over (m+1)} {_1F_1}\left(1,{m+3\over 2},x^2\right)
\label{element definitions}
\ea
and where $\gamma$ and $_1F_1$ respectively denote the incomplete gamma function and
confluent hypergeometric function. 

There are some properties of $\wdm$ worthy of note. Firstly, as 
is clear from the form of (\ref{well wave function}),
$\vert\psi^H_0(x_1,x_2,...,x_N)\vert^2$ is precisely the probability density function for
the distribution of eigenvalues of the Gaussian Unitary Ensemble (GUE) of
random matrix theory, and hence $\wdm$ can be interpreted as a certain
average over the GUE\cite{cmp}. Secondly it is obvious that since the
ground states of the
noninteracting Fermi and impenetrable Bose systems differ only in the
presence of the absolute value in the wave function of the latter, we
see by comparing (\ref{well element integral}) with (\ref{well fermi
  bose integral comparison}) that the density matrix for a harmonically trapped
noninteracting Fermi system is given by (\ref{well dm}) with $\elem$ replaced by
$f_{j,k}(x,y)$. From this observation it is immediately clear that the
densities for the Fermi and Bose cases are identical since on setting
$x=y$ in (\ref{well fermi bose comparison}) we see that
$b_{j,k}(x,x)=f_{j,k}(x,x)$. The global large $N$ limit of the density in this
case results in the Wigner semi circle
law\cite{baker,johanssonfluctuations} 
\be
\rho_N^H(x,x)={\w\over \pi}\sqrt{1-{x^2\over 2N}},\quad \vert x\vert\le\w ,
\ee
while for $\vert x\vert>\sqrt{2N}$ we have $\rho_N^H(x,x)=0$ to
leading order.
An oscillatory correction to this leading global asymptotic form has
been given in \cite{kalisch}.
Finally we note from the
second equality in (\ref{element definitions}) that since the $_1F_1$
function is entire,  $\wdm$ is
analytic everywhere in the finite $(x,y)$ plane except along the
diagonal $y=x$, where it has discontinuities in its first
derivatives. Such observations are important in choosing a numerical quadrature method. 
 
\subsubsection{Occupation numbers}
Quite generally, for a many body quantum system the eigenvalue equation 
\be
\int\rho_N(x,y) \phi_j(y)dy=\lambda_j\phi_j(x), \qquad j=0,1,2,... 
\label{integral equation}
\ee
defines the natural orbitals, $\phi_j(x)$, which
have the physical
interpretation of being effective single
particle states, and the eigenvalues $\lambda_j$ which are interpreted
as the occupation numbers for these natural orbitals. When the system lies on a circle the periodicity
implies that the natural orbitals are simply plane waves, and so the
eigenvalues are given by the Fourier coefficients of $\rho_N(x-y)$ and
hence the momentum distribution and the set of natural orbital
occupations coincide. In the general case, in which
$\rho_N(x,y)$ is not translationally invariant, this is no longer
true. Recent work \cite{girardeau8,papenbrock} has focused attention on the
computation and analysis of the momentum distribution for harmonically trapped
impenetrable bosons. A more demanding task is the investigation  of the $N$
dependence of the occupation of the lowest natural orbital, $\lambda_0$,
the fundamental quantity of interest in discussing BEC-like
coherence. We undertake a
numerical analysis of $\lambda_0$ in Section \ref{numerical
investigation of the eigenvalues}, and then in
Section \ref{occupation numbers and natural orbitals as n to infty} we
go on to discuss how a new analytic result, obtained in Section \ref{peter}, 
yields its large $N$ scaling. 

\subsection{Numerical investigation of the eigenvalues of $\wdm$}
\label{numerical investigation of the eigenvalues}
The integral equation (\ref{integral equation}) was solved numerically for  $2\le N\le30$ to obtain the
corresponding values of $\lambda_0(N)$ and $\lambda_1(N)$.
In the present section we carefully analyse this numerical
data in order to determine the $N$ dependence of $\lambda_0(N)$ and $\lambda_1(N)$. Before getting under way however, a few comments are in
order.
\subsubsection{Perfunctory remarks}
Since our ultimate goal is to infer from the sequence of
computed $\lambda_0(N)$ and $\lambda_1(N)$ values how these quantities grow
with increasing $N$, it is not only the accuracy  and number of the computed
$\lambda$ values that are obtained that is important, but equally
important is how we infer the large $N$ behaviour from
these computed values. Naive approaches such as fitting the data to a log/log plot
in order to obtain the exponent in $\lambda_0\sim N^{\alpha}$
\cite{girardeau10} prove to be of no use. A striking demonstration of this
fact occurs if we attempt to apply this procedure to our numerical $c_0(N)$
data for the system on a circle. In this case we know a priori by
analytic arguments that the exponent is precisely $1/2$. However, fitting 
the $c_0(N)$ data with $2\le N\le 10$ via a log/log plot predicts an exponent
of $\alpha=0.64$, and even using  $2\le N\le 70$ only
forces the exponent to drop to $\alpha=0.60$. A similar analysis 
 for the harmonically trapped system produces analogous results. Clearly such an approach
will only be useful for very large $N$, and more suitable methods of
analysis are required. 

The method we shall use to analyse the $N$ dependence of the computed $\lambda_0$
and  $\lambda_1$ values is motivated by series analysis techniques. We 
assume an ansatz for $\lambda_0(N)$ of the form 
\be
\lambda_0(N)=
aN^p + b +{c\over N^x},
\label{series analysis ansatz}
\ee
and similarly for $\lambda_1(N)$. We then fit our numerical data to this
ansatz, varying the parameters $p$ and $x$ and seeking to minimise the
$\chi^2$ value for the fit. We have already noted that the fits of the numerical data for the system on
the circle to this form of ansatz are in very good agreement with the
known analytic results. We shall see in Section \ref{occupation numbers and natural orbitals as n to infty}
that again our numerical
fits to this ansatz agree remarkably with  analytic
results that we obtain via independent means. 
\subsubsection{Numerical Results}
\label{well numerical results}
To numerically solve 
\be
\int_{-\infty}^{\infty}\rho_N^H(x,y) \phi_j(y)dy=\lambda_j\phi_j(x), \qquad j=0,1,2,... 
\label{well integral equation}
\ee
we used the so called
quadrature method \cite{delves} which replaces the integral equation 
with an approximate matrix equation. Specifically, we factored out $e^{-y^2/2}$
in (\ref{well dm}) and applied a  Gauss-Hermite quadrature (with
abscissae $\xi_i$, weights $w_i$ and $1\le i\le Z$) to
$e^{y^2/2}\wdm \phi(y)$. Setting $x=\xi_i$ in (\ref{well integral
  equation}) for each $i$, we thus obtain $Z$ linear equations,
 which then defines a matrix approximation to our integral
operator. This matrix, $S$, can be so chosen that
\ba
S_{l,m} 
&=&
e^{-x^2_l/4 -x^2_m/4}\sqrt{w_l w_m}{2^{N-1}\over \sqrt{\pi}\Gamma(N)}
\det\left[{2^{(j+k)/2}\over 2\sqrt{\pi}\sqrt{\Gamma(j)\Gamma(k)}}b_{j,k}(\xi_l,\xi_m)\right]_{j,k=1,...N-1}
\nonumber \\
\label{matrix}
&&l,m=1,2,...Z,
\ea 
where a suitable transformation has been used to force $S$ to be symmetric, since
such a transformation 
makes the numerical computations more efficient. We computed
the required $b_{j,k}(\xi_l,\xi_m)$ using the representation in terms of incomplete
gamma functions (\ref{well fermi bose comparison}) with (\ref{element definitions}).

%% Our task of evaluating the occupation numbers for a system of $N$
%% harmonically trapped impenetrable bosons is now seen to be the task of
%% numerically determining the eigenvalues of (\ref{matrix}). 
Due to the
complexity of $S$, each of its elements being given in terms of an $(N-1)$ dimensional 
determinant, we find that as $Z$ and $N$ are increased round off error becomes an
important obstacle. For $N=30$ convergence was down
to two significant figures and by $N=35$ we found that $64$ bit precision became
insufficient and to obtain higher values of $N$ a multiprecision
routine would be necessary. As $N$ approached $N=30$ the fits of the
data to (\ref{series analysis ansatz})
became rather sensitive to the precision of the $\lambda_j$ used. Due to this we chose to fit only
the data $2\le N\le 27$ to the ansatz (\ref{series analysis ansatz})
since for this set we could be sure that the precision was at least three
significant figures (actually considerably more for the lower $N$
values). With the computed $\lambda_j$ analysed according to
(\ref{series analysis ansatz}) however, it turns out that a good deal
of large $N$ asymptotic information is already present even at these
relatively low values of $N$. This will become evident in Section \ref{occupation numbers and natural orbitals as n to infty}.

%% As we will see in Section \ref{numerical solution of the
%%   scaled large N integral equation}, provided the data is analysed
%% carefully it is possible to extract a good deal of 
%% As we will see in Section \ref{numerical solution of the
%%   scaled large N integral equation} however, a good deal of large $N$
%% asymptotic information is already present even at such
%% relatively low $N$ values, and can be extracted provided the data is analysed appropriately.

Systematically varying the parameters $p$ and $x$ in (\ref{series
  analysis ansatz}), we observed a dramatic dip of three to four
  orders of magnitude in the $\chi^2$ of the fit when $p$
hit the critical value of $p=0.5$. For $p=0.45$ and $p=0.55$ the best
$\chi^2$ values as $x$ was varied were of the order $10^{-3}$, and
these 
$\chi^2$ values became larger as $p$ was moved away from $p=0.5$ on either side. At
$p=0.5$ however, the best $\chi^2$ value was of the order $10^{-7}$,
which occurred 
for $x$ in the neighbourhood of $2/3$. This
very steep dip in $\chi^2$  precisely in the neighbourhood of $p=0.5$
  indicates
that the exponent of the leading order behaviour of $\lambda_0$ with $N$
is in fact $1/2$, just as is the case on the circle. A similar analysis
performed for $\lambda_1$ yielded the same sharp dip in $\chi^2$ at 
$p=0.5$, again dropping from around $10^{-3}$ on either side of $p=0.5$
to around $10^{-7}$ at $p=0.5$. The coefficients of the correction terms
for $\lambda_1$ turn out to be identical, to two decimal places, to
those for $\lambda_0$, however it now appears that the best value of
$x$ is in the neighbourhood of $x=4/3$. Specifically, fitting the
computed $\lambda_0$ and $\lambda_1$ values with $2\le N\le 27$
we obtained
\be
\lambda_0(N)=1.43\sqrt{N} -0.56 +{0.12\over N^{2/3}}
\label{e0fit}
\ee

\be
\lambda_1(N)=0.61\sqrt{N} -0.56 +{0.12\over N^{4/3}}.
\label{e1fit}
\ee
with $\chi^2\sim 10^{-7}$. 
Again we remark that the similarity of the constant in these fits with that for the case of
the uniform system, (\ref{c_0numerical}) and (\ref{c_1numerical}), is quite suggestive. 

\begin{figure}
\includegraphics{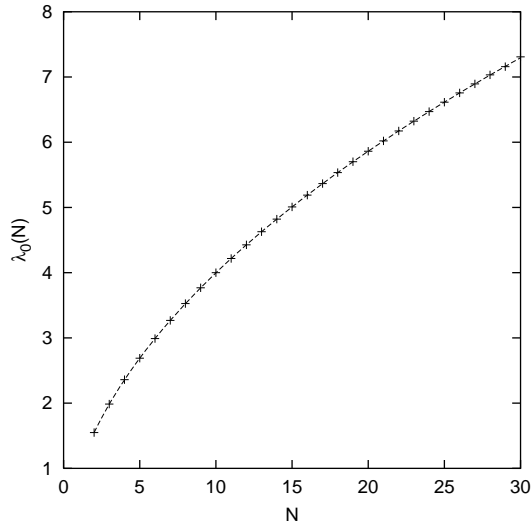}
\caption{Comparison of the computed values of $\lambda_0(N)$, with the
  fit of this data to the series ansatz, (\ref{e0fit}).} 
\label{e0}   
\end{figure}

\begin{figure}
\includegraphics{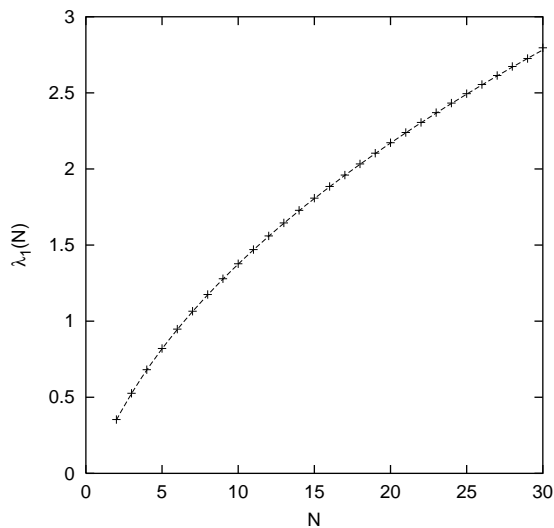}
\caption{Comparison of the computed values of $\lambda_1(N)$, with the fit of this data to the series ansatz, (\ref{e0fit}).} 
\label{e1}   
\end{figure}

In FIGS.~\ref{e0},\ref{e1}  we plot the computed numerical values for $\lambda_0(N)$ and
$\lambda_1(N)$ for $2\le N\le 30$ together with the fits
(\ref{e0fit}),(\ref{e1fit}).
%% The fits are very
%% sensitive to the precision of the computed values of $\lambda_k$, and
%% although quite stable with regard to varying the largest value of $N$
%% included in the fits with $N\le 27$, including only three more
%% points from $N=28,29,30$ in which the precision was slightly less
%% certain resulted in increased $\chi^2$ values for all values of $p$.

Recently Papenbrock \cite{papenbrock} discussed the computation of the density
matrix using a form similar to (\ref{well dm}). 
%% identical to (\ref{well dm}) except
%% that the elements $\elem$ were integrated numerically rather than
%% computed by identifying (\ref{well
%%   fermi bose comparison}). 
Information regarding the leading $N$ behaviour of
$\lambda_0$ was inferred indirectly from a numerical investigation of
the momentum distribution, by way of an interesting scaling argument, rather than by analysing the $\lambda_0(N)$
directly as we have done. This argument predicts that $\lambda_0\propto
\sqrt{N}$ for large $N$, in agreement with our analytical study to be
presented below.

In the next Section we construct the large $N$ asymptotic form for
$\wdm$. 
This result allows the $\sqrt{N}$ scaling of $\lambda_j$
to be put on a firm analytic footing as a straightforward
corollary. 
\subsection{Global asymptotics from a log-gas picture}
\label{peter}
We would like to derive the analogue of the asymptotic formula
(\ref{large x expansion}) for the impenetrable Bose gas in a harmonic
trap. Now the latter is given as a determinant by (\ref{well dm}). To our knowledge unlike the situation for the Toeplitz
determinant (\ref{toeplitz}), there is no known general theorem giving
the asymptotic form of determinants  of the type  (\ref{well dm}),
or equivalently multidimensional integrals of the form
(\ref{well multidimensional integral}). (If each factor $\vert x-x_l\vert \vert y-x_l\vert$ in
(\ref{well multidimensional integral}) is replaced by $e^{-g(x_l)}$ for $g$ analytic, such
general asymptotics are known \cite{johanssoncompactclassicalgroups}). However, as we will now demonstrate, a Coulomb gas
argument which can be used to give a heuristic derivation of
(\ref{large x expansion}) does carry over to the case of the density
matrix for the impenetrable Bose gas in a harmonic trap, thus allowing
the global asymptotics of the latter to be predicted.

Let us first revise the derivation of the asymptotic form (\ref{large
  x expansion}), slightly modified relative to the presentation in
  \cite{peter2}    
so it is best adapted to application in the harmonic trap case. The
  starting point is the multidimensional integral formula
  (\ref{intermediate toeplitz}) for $\dm$. Put
\ba
Z_n^C\left((\phi_1,q_1),(\phi_2, q_2)\right)
&:=& 
\vert e^{i\phi_1}-e^{i\phi_2}\vert^{2q_1 q_2}
\nonumber \\
&\times&
\int_{0}^{2\pi}d\theta_1 ...\int_{0}^{2\pi}d\theta_n \prod_{l=1}^{n}
\vert e^{i\phi_1}-e^{i\theta_l}\vert^{2q_1}
\vert e^{i\phi_2}-e^{i\theta_l}\vert^{2q_2}
\nonumber \\
&\times&
\prod_{1\le j<k\le n}
\vert e^{i\theta_k}-e^{i\theta_j}\vert^2
\ea
This is the classical configuration integral for $n$ mobile unit charges, and two fixed impurity charges (at angles $\phi_1$, $\phi_2$ and of
charge $q_1$, $q_2$) on the unit circle interacting via the
logarithmic potential 
$- q_a q_b \log\vert e^{i\theta_a}-e^{i\theta_b}\vert$ 
at inverse temperature $\beta=2$. In terms of this notation, and that of (\ref{circle notation}), we can write 

\be
\dm = {2\pi N\over \vert 1-e^{2it}\vert^{1/2}}
{Z^C_{N-1}\left((0,1/2),(2t,1/2)\right)\over Z^C_N}
\label{n2}
\ee
where 
\be
Z^C_N:=Z_N^C((\,\cdot\,,0),(\,\cdot\,,0))=(2\pi)^N N!
\label{n3}
\ee
Now for $N$ large and $t$ fixed the impurity charges are effectively
separated by a macroscopic amount (in units of the mean interparticle
separation) so we expect the factorization
\ba
{Z^C_{N-1}\left((0,1/2),(2t,1/2)\right)\over Z_N^C}
&\sim&
{Z^C_{N-1}((0,1/2))\over Z^C_{N-1/2}}
{Z^C_{N-1}((2t,1/2))\over Z^C_{N-1/2}}
\nonumber \\
&=&
\left({Z^C_{N-1}((0,1/2))\over Z^C_{N-1/2}}\right)^2
\label{n4}
\ea

Now the reason for $N-1/2$ in the classical configuration integral of
the denominators is to exactly balance the total charge in the
numerator, as in the original expression (\ref{n2}). Of course to
evaluate $Z^C_{N-1/2}$ we must analytically continue off the integers
but this is immediate from the exact evaluation in (\ref{n3}).
Furthermore, we know from (\ref{intermediate toeplitz}) that 
\ba
Z^C_{N-1}((0,1/2)
&=&
(2\pi)^{N-1} \left({G(N+1) G(3/2)\over G(N+1/2)}\right)^2
\nonumber \\
&\sim&
(2\pi)^{N-1/2} N^{-1/4} e^{-N} N^N (G(3/2))^2
\label{n5}
\ea
while it follows from (\ref{n3}) and Stirling's formula that 
\be
Z^C_{N-1/2}=(2\pi)^{N-1/2}\Gamma(N+1/2)\sim (2\pi)^N N^N e^{-N}
\label{n6}
\ee
and so 
\be
{Z_{N-1}((0,1/2))\over Z_N}\sim {(G(3/2))^2\over \sqrt{2\pi}N^{1/4}}.
\label{n7}
\ee
Substituting (\ref{n7}) and (\ref{n5}) in (\ref{n4}) then substituting
the result in (\ref{n2}) reclaims (\ref{large x expansion}). 

This  log-gas argument readily generalises to provide the global
asymptotics for
\be
\varrho^H(X,Y):=(2N)^{1/2}\rho_N(\sqrt{2N}X, \sqrt{2N}Y)
\label{n8}
\ee 
with $-1<X,Y<1$ fixed and $N\to\infty$. Let us write 
\ba
Z^H_{n,m}((X_1,q_1),(X_2,q_2))
&:=&
\vert X_1 - X_2\vert^{2 q_1 q_2}e^{-mq_1 X_1^2/2}e^{-mq_2 X_2^2/2}
\nonumber \\
&\times&
\int_{-2}^{2}dx_1 ...\int_{-2}^{2}dx_n 
\prod_{l=1}^n 
e^{-mx_l^2/2}\vert X_1 -x_l\vert^{2q_1}\vert X_2 - x_l \vert^{2q_2}
\nonumber \\
&\times&
\prod_{1\le j<k\le n}
\vert x_k -x_j\vert ^2
\ea
where $m=n+q_1 +q_2$. This is the classical configuration integral for
$n$ mobile unit charges and two fixed impurity charges (of charge
$q_1$,$q_2$ at $X_1$,$X_2$ respectively) interacting on the interval
$[-2,2]$ via the logarithmic potential $-q_a q_b \log\vert x_a
-x_b\vert$ at inverse temperature $\beta=2$. The charges are also
subject to the one body potential $(mq_a/4) x_a^2$. (We work in the
scaled interval [-2,2] as this is the convention used in
\cite{brezin}, the results of which we will be using subsequently.)
Now it is well known from the random matrix interpretation of
$\vert\psi_N^H\vert^2$ (recall \cite{cmp}) that the support for the
density is the interval $[-\sqrt{2N},\sqrt{2N}]$. Doing this then
changing variables shows 
\be
\varrho(X,Y) = {2N\over
  \vert2X-2Y\vert^{1/2}}{Z^H_{N-1,M}((2X,1/2),(2Y,1/2))\over Z^H_{N,N}}
\label{n10}
\ee
where
\[
Z^H_{n,m}=Z^H_{n,m}((\,\cdot\,,0),(\,\cdot\,,0)).
\] 

Now analogous to (\ref{n4}), for $N$ large and $-2<X,Y<2$ fixed we
expect the factorisation
\be
{Z^H_{N-1,N}((2X,1/2)(2Y,1/2))\over Z^H_{N,N}}\sim
{Z^H_{N-1,N-1/2}((2X,1/2))\over Z^H_{N-1/2,N-1/2}}
{Z^H_{N-1,N-1/2}((2Y,1/2))\over Z^H_{N-1/2,N-1/2}}.
\label{n11}
\ee
But in the notation of Brezin and Hikami (Eq. (46) of \cite{brezin})
\be
{Z^H_{N-1,N-1/2}((2X,1/2))\over
  Z^H_{N-1/2,N-1/2}}={(N-1/2)!\over N!}M_1(2X)
\label{n12}
\ee
and it is shown in \cite{brezin} (see also \cite{strahov}) that 
\be
M_1(2X)\sim(2\pi)^{-1/2}(2N(1-X^2)^{1/2})^{1/4}(G(3/2))^2
\label{n13}
\ee
where we have used the fact \cite{peter2} 
\[
\prod_{l=0}^{K-1}
{l!\over (K+l)!}
={(G(K+1))^2\over G(2K+1)}
\]
Substituting (\ref{n13}) in (\ref{n12}) and noting that
$(N-1/2)!/N!\sim N^{-1/2}$, we then obtain from (\ref{n11}) and
(\ref{n10}) the sought asymptotic formula
\be
\varrho^H(X,Y)\sim N^{1/2}{(G(3/2))^4\over
  \pi}{(1-X^2)^{1/8}(1-Y^2)^{1/8}\over\vert X-Y\vert^{1/2}}
\label{n14}
\ee 
\subsection{Occupation numbers and natural orbitals as $N\to\infty$}
\label{occupation numbers and natural orbitals as n to infty}
Using (\ref{n14}) we can determine the large $N$ behaviour of the
occupation numbers and natural orbitals.
Taking the large $N$ limit of the integral equation (\ref{well integral equation}), utilising
(\ref{n14}) and (\ref{n8}) we obtain
\be
 {(G(3/2))^4\over \pi}  
\int_{-1}^{1}dX\;
  {(1-X^2)^{1/8}(1-Y^2)^{1/8}\over\vert X-Y\vert^{1/2}}
  \varphi_j(X)
={\lambda_j\over \sqrt{N}}\varphi_j(Y),
\label{scaled integral equation}
\ee
where we have defined the scaled limiting large $N$ natural orbitals 
\be
(2N)^{1/4}\phi_j(x)\to\varphi_j(X).
\label{scaled phi}
\ee
This scaling for $\phi_j(x)$ arises since for large $N$ the
density matrix $\wdm$
has support $[-\w,\w]$, and so the integral equation (\ref{well integral
  equation}) tells us that $\phi_j(x)$ must also, hence
normalising  $\phi_j(x)$ on this interval implies (\ref{scaled phi}).

From (\ref{scaled integral equation}), the large $N$ scaling of
$\lambda_j$ is immediate. Since both the left hand side of (\ref{scaled integral equation}) and
$\varphi_j(Y)$ are independent of $N$, we must have that for large $N$
\be
\lambda_j \propto \sqrt{N}.
\label{scaled lambda}
\ee 
We note that this argument has regarded $j$ as a fixed parameter. For $j\gg N$ we
expect the behaviour of $\lambda_j$ to be similar to that of  
$c_j(N)$  given by (\ref{large n}), since very excited states should not feel the presence of
the trap.

Hence from a knowledge of the asymptotic behaviour of $\wdm$ we have
demonstrated not only that $\lambda_0\sim
\sqrt{N}$, but that the other low lying modes must also display this $N$ dependence. We
can go one step further however and numerically solve 
(\ref{scaled integral equation}) to obtain the proportionality
constants missing from (\ref{scaled lambda}).

Factoring out the exact constants, (\ref{scaled integral equation})
becomes
 \be
 \int_{-1}^{1}dX\; {\varphi_j(X) (1-X^2)^{1/8}\over \vert X-
 Y\vert^{1/2}}
 ={\bar \lambda_j}{\varphi_j(Y)\over (1-Y^2)^{1/8}},
 \label{norms integral equation}
 \ee
 where 
 \be
 {\lambda_j}= {[G(3/2)]^4\over \pi}{{\bar\lambda_j}\sqrt{N}}.
 \label{jeff}
 \ee
We solved (\ref{norms integral equation}) using a similar method to that discussed in
Section \ref{well numerical results}, to obtain 
\be
{\bar \lambda_0}= 3.438...\qquad
{\bar \lambda_1}= 1.47...\qquad
{\bar \lambda_2}= 1.13...
\label{tdl evalues}
\ee
and hence, using $[G(3/2)]^4/\pi= 0.4160...$ we find that in the large $N$ limit 
\ba
\lambda_0&=& 1.430\sqrt{N},
\label{norm fits 0}\\
\lambda_1&=& 0.61 \sqrt{N},
\label{norm fits 1}\\
\lambda_2&=& 0.47 \sqrt{N}.
\ea
The results (\ref{norm fits 0}) and (\ref{norm fits 1}) are to be
compared with the numerical fits (\ref{e0fit}) and (\ref{e1fit}) for
$\lambda_0$ and  $\lambda_1$. The agreement is remarkable. 

\begin{figure}
\includegraphics{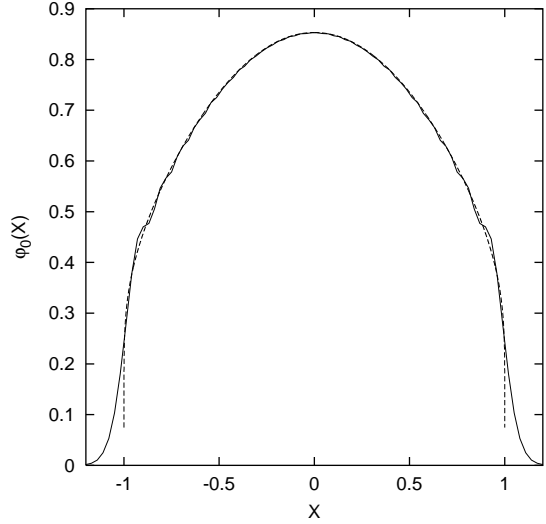}
\caption{Comparison of the limiting large $N$ natural orbital
  $\varphi_0(X)$, denoted
  by the dashed line, with the
  the scaled natural orbital $\w \phi_0(x)$ for $N=25$.} 
\label{log0}   
\end{figure}

\begin{figure}
\includegraphics{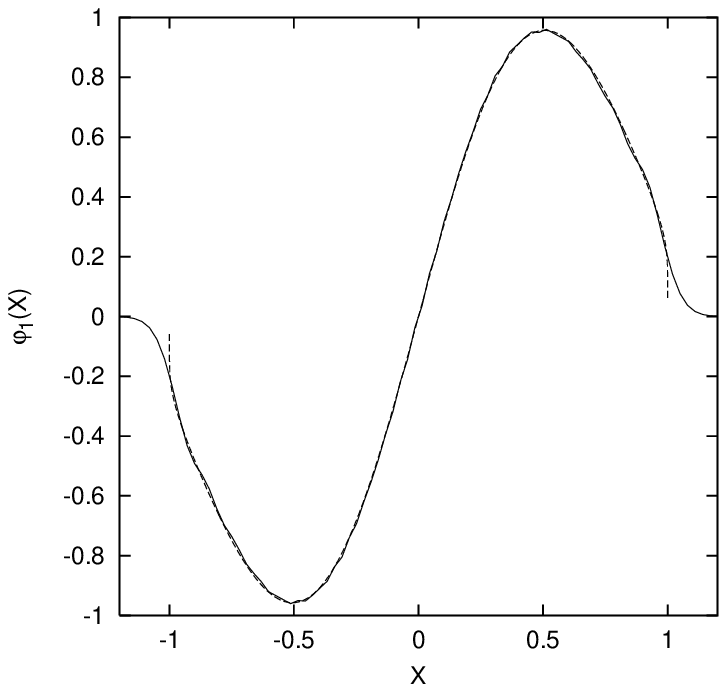}
\caption{Comparison of the limiting large $N$ natural orbital
  $\varphi_1(X)$, denoted
  by the dashed line, with the
  the scaled natural orbital $\w \phi_1(x)$ for $N=25$.} 
\label{log1}   
\end{figure}

\begin{figure}
\includegraphics{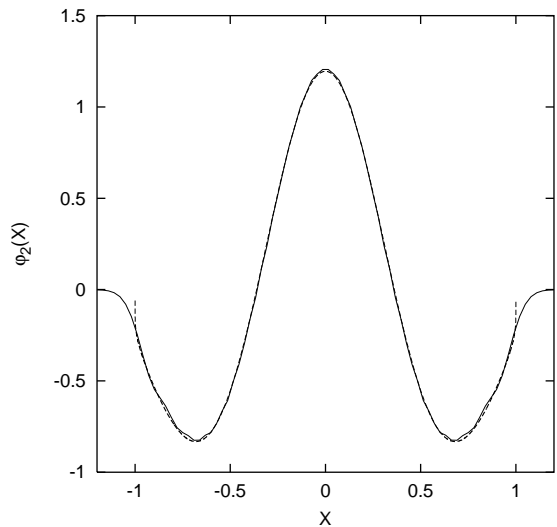}
\caption{Comparison of the limiting large $N$ natural orbital
  $\varphi_2(X)$, denoted
  by the dashed line, with the
  the scaled natural orbital $\w \phi_2(x)$ for $N=25$.} 
\label{log2}   
\end{figure}

The limiting scaled natural orbitals are also of interest. FIGS.~\ref{log0},\ref{log1},\ref{log2} show
$\varphi_j(X)$, together with
$(\w)^{1/4}\phi_j(x)$ computed from (\ref{well integral equation}) with
$N=25$, for $k=0,1,2$. It is clear that already
by $N=25$ the eigenfunctions of the large $N$ equation (\ref{scaled
  integral equation}) provide a good approximation to the finite
$N$ results, only really differing outside the support. The cause of
the sharp decrease as $X\to \pm 1$ can be understood by noting that
(\ref{norms integral equation}) implies that 
$\varphi_j(X)$ vanishes like $(1-X^2)^{1/8}$ in this limit.

We remark that $\varphi_j(X)$ is an even (odd) function when $j$ is even
(odd). 
For reference, we list the computed values of the
central maxima and the area under the curve for $\varphi_0(X)$ and $\varphi_2(X)$. Whilst the area under $\varphi_2(X)$ is certainly smaller
than that under $\varphi_0(X)$, it is not so much smaller as to be negligible.
\be
\varphi_0(0)=0.853... \qquad, \varphi_2(0)=1.20...
\ee 
\be
\int_{-1}^1dX\;\varphi_0(X)=1.38...\qquad
\int_{-1}^1dX\;\varphi_2(X)=0.21...
\ee 

\section{Conclusion}
\label{conclusion}
We have given an exhaustive study of the particle occupation numbers
for the $1d$ impenetrable boson systems. Our feature results are presented
in Section \ref{impenetrable bosons trapped in a harmonic well} for the  harmonically trapped systems. These results
are of such  strikingly similarity to those for the untrapped boson systems
presented in Section \ref{impenetrable bosons on a circle} as to make all the results collectively invaluable.

We  commend the challenge to the experimentalists of measuring these
signature quantities.  
\begin{acknowledgments}
The authors would like to thank the Australian Research Council for
funding this work.
\end{acknowledgments}
\appendix
\section{Summary of ground state energies of systems 
{subject} to
various boundary conditions }
\label{energy appendix}
We list here for ease of reference the ground state
energies per particle, $U_0/N$, and the Fermi energies,
$\epsilon_F$, for finite one dimensional systems of
impenetrable bosons of mass $m$ subject to periodic, Dirichlet, and
Neumann boundary conditions and the analogous results in the
thermodynamic limit, as well as for impenetrable bosons confined
by a harmonic well. These results are identical to the analogous noninteracting
Fermi results.
\begin{table*}[ht]
\begin{tabular}{|c|c|c|c|} \hline\hline
System &$U_0/N$ &$\epsilon_F$  \\
\hline\hline
Circle                    &$(N^2-1)\pi^2\hbar^2/6mL^2$& $(N-1)^2\pi^2\hbar^2/2mL^2$ \\
\hline
Dirichlet            &$(N+1)(N+1/2)\pi^2\hbar^2/6mL^2$& $N^2    \pi^2\hbar^2/2mL^2$ \\
\hline
Neumann              &$(N-1)(N+1/2)\pi^2\hbar^2/6mL^2$& $(N-1)^2\pi^2\hbar^2/2mL^2$ \\
\hline
Thermodynamic limit  &$N^2\pi^2\hbar^2/6mL^2$         & $N^2    \pi^2\hbar^2/2mL^2$ \\
\hline
Harmonic well, $V(z)=m\omega_z^2z^2/2$       &$N\hbar\omega_z/2$& $(N-1/2)\hbar\omega_z$  \\
\hline\hline
\end{tabular}
\caption{Summary of ground state energies per particle and Fermi
energies for a systems of impenetrable bosons subject to periodic,
Dirichlet and Neumann boundary conditions, and a harmonic well. }
\label{appendixtable}
\end{table*}

\section{Trial series solution for $\norm$}
\label{norm appendix}
We demonstrate here that it is possible to learn a good deal about $\norm$ by suitably approximating the
integrand of (\ref{norms integral equation}). To begin with, simply
taking $Y=0$ in (\ref{norms integral equation}) 
yields the following bound for $\norm$ 
\ba
\norm &=&
\int_{-1}^1{\varphi_0(X)\over\varphi_0(0)}{(1-x^2)^{1/8}\over\vert x\vert^{1/2}}dx
\nonumber \\
\norm &<&
2\int_0^1{(1-x^2)^{1/8}\over\vert x\vert^{1/2}}dx= B(1/4,9/8)=3.84108...
\ea

A very good approximation to the value of
$\norm$  can be obtained by inserting the formal Maclaurin expansion for
$\varphi_0(X)$,
\be
\varphi_0(X)=\varphi_0(0)(1-AX^2-BX^4-...)
\label{maclaurin}
\ee
into (\ref{norms integral equation}) and equating
coefficients of powers of $Y$ to obtain the unknown coefficients and
$\norm$. Keeping only terms to order $Y^2$ and $X^2$ in the expansion
of the integrand we thus obtain 
\be
\int_0^{1+Y}{dt\over \sqrt{t}}[1-(A+{1\over
  8})(t-Y)^2]+\int_0^{1-Y}{dt\over \sqrt{t}}[1-(A+{1\over
  8})(t+Y)^2]=\norm[1-(A-{1\over 8})Y^2] 
\ee
which then yields, upon equating coefficients
\ba
\norm
&=&
4-{4\over 5}\left[A+{1\over 8}\right],
\\
\norm \left[A-{1\over 8}\right]
&=&
{1\over 2}+{3\over 2}\left[A+{1\over 8}\right],
\ea 
and the solution of these two simultaneous equations gives
\be
A=0.57627... \qquad \norm=3.43899...
\ee

Carrying out this procedure again now keeping terms to fourth order we
find that $A$ and $\norm$ shift only slightly
\be
A=0.57682... \qquad B=0.01914...
\ee
\be
\norm=3.4378...
\ee
which shows that the procedure is quite stable.
This value of  $\norm$ is in excellent agreement with our
numerical solution of (\ref{norms integral equation}) given in
(\ref{tdl evalues}).

%% \bibliography{prabib}

\begin{thebibliography}{45}
\expandafter\ifx\csname natexlab\endcsname\relax\def\natexlab#1{#1}\fi
\expandafter\ifx\csname bibnamefont\endcsname\relax
  \def\bibnamefont#1{#1}\fi
\expandafter\ifx\csname bibfnamefont\endcsname\relax
  \def\bibfnamefont#1{#1}\fi
\expandafter\ifx\csname citenamefont\endcsname\relax
  \def\citenamefont#1{#1}\fi
\expandafter\ifx\csname url\endcsname\relax
  \def\url#1{\texttt{#1}}\fi
\expandafter\ifx\csname urlprefix\endcsname\relax\def\urlprefix{URL }\fi
\providecommand{\bibinfo}[2]{#2}
\providecommand{\eprint}[2][]{\url{#2}}

\bibitem[{\citenamefont{Cornell and Wieman}(2002)}]{cornell}
\bibinfo{author}{\bibfnamefont{E.~A.} \bibnamefont{Cornell}} \bibnamefont{and}
  \bibinfo{author}{\bibfnamefont{C.~E.} \bibnamefont{Wieman}},
  \bibinfo{journal}{Rev. Mod. Phys.} \textbf{\bibinfo{volume}{74}},
  \bibinfo{pages}{875} (\bibinfo{year}{2002}).

\bibitem[{\citenamefont{Ketterle}(1999)}]{ketterle}
\bibinfo{author}{\bibfnamefont{W.}~\bibnamefont{Ketterle}},
  \bibinfo{journal}{Physics Today} \textbf{\bibinfo{volume}{52}},
  \bibinfo{pages}{30} (\bibinfo{year}{1999}).

\bibitem[{\citenamefont{Pritchard}(2002)}]{pritchard}
\bibinfo{author}{\bibfnamefont{D.~E.} \bibnamefont{Pritchard}},
  \bibinfo{journal}{MIT Physics Annual} \textbf{\bibinfo{volume}{15}},
  \bibinfo{pages}{40} (\bibinfo{year}{2002}).

\bibitem[{\citenamefont{Olshanii}(1998)}]{olshanii}
\bibinfo{author}{\bibfnamefont{M.}~\bibnamefont{Olshanii}},
  \bibinfo{journal}{Phys. Rev. Lett.} \textbf{\bibinfo{volume}{81}},
  \bibinfo{pages}{938} (\bibinfo{year}{1998}).

\bibitem[{\citenamefont{Dettmer et~al.}(2001)\citenamefont{Dettmer, Hellweg,
  Ryytty, Arlt, Ertmer, Sengstock, Petrov, Shlyapnikov, Kreutzmann, Santos
  and  Lewenstein}}]{dettmer}
\bibinfo{author}{\bibfnamefont{S.}~\bibnamefont{Dettmer}},
  \bibinfo{author}{\bibfnamefont{D.}~\bibnamefont{Hellweg}},
  \bibinfo{author}{\bibfnamefont{P.}~\bibnamefont{Ryytty}},
  \bibinfo{author}{\bibfnamefont{J.~J.} \bibnamefont{Arlt}},
  \bibinfo{author}{\bibfnamefont{W.}~\bibnamefont{Ertmer}},
  \bibinfo{author}{\bibfnamefont{K.}~\bibnamefont{Sengstock}},
  \bibinfo{author}{\bibfnamefont{D.~S.} \bibnamefont{Petrov}},
  \bibinfo{author}{\bibfnamefont{G.~V.} \bibnamefont{Shlyapnikov}},
  \bibinfo{author}{\bibfnamefont{H.}~\bibnamefont{Kreutzmann}},
  \bibinfo{author}{\bibfnamefont{L.}~\bibnamefont{Santos}} \bibnamefont{and}
  \bibinfo{author}{\bibfnamefont{M.}~\bibnamefont{Lewenstein}}, 
  \bibinfo{journal}{Phys. Rev. Lett.}
  \textbf{\bibinfo{volume}{87}}, \bibinfo{pages}{160406}
  (\bibinfo{year}{2001}).

\bibitem[{\citenamefont{G\"orlitz et~al.}(2001)\citenamefont{G\"orlitz, Vogels,
  Leanhardt, Raman, Gustavson, Abo-Shaeer, Chikkatur, Gupta, Inouye, Rosenband
  and Ketterle}}]{gorlitz}
\bibinfo{author}{\bibfnamefont{A.}~\bibnamefont{G\"orlitz}},
  \bibinfo{author}{\bibfnamefont{J.~M.} \bibnamefont{Vogels}},
  \bibinfo{author}{\bibfnamefont{A.~E.} \bibnamefont{Leanhardt}},
  \bibinfo{author}{\bibfnamefont{C.}~\bibnamefont{Raman}},
  \bibinfo{author}{\bibfnamefont{T.~L.} \bibnamefont{Gustavson}},
  \bibinfo{author}{\bibfnamefont{J.~R.} \bibnamefont{Abo-Shaeer}},
  \bibinfo{author}{\bibfnamefont{A.~P.} \bibnamefont{Chikkatur}},
  \bibinfo{author}{\bibfnamefont{S.}~\bibnamefont{Gupta}},
  \bibinfo{author}{\bibfnamefont{S.}~\bibnamefont{Inouye}},
  \bibinfo{author}{\bibfnamefont{T.}~\bibnamefont{Rosenband}} \bibnamefont{and}
  \bibinfo{author}{\bibnamefont{W.}~\bibnamefont{Ketterle}}, 
  \bibinfo{journal}{Phys. Rev. Lett.}
  \textbf{\bibinfo{volume}{87}}, \bibinfo{pages}{130402}
  (\bibinfo{year}{2001}).

\bibitem[{\citenamefont{Greiner et~al.}(2001)\citenamefont{Greiner, Bloch,
  Mandel, H\"ansch, and Esslinger}}]{greiner}
\bibinfo{author}{\bibfnamefont{M.}~\bibnamefont{Greiner}},
  \bibinfo{author}{\bibfnamefont{I.}~\bibnamefont{Bloch}},
  \bibinfo{author}{\bibfnamefont{O.}~\bibnamefont{Mandel}},
  \bibinfo{author}{\bibfnamefont{T.~W.} \bibnamefont{H\"ansch}},
  \bibnamefont{and}
  \bibinfo{author}{\bibfnamefont{T.}~\bibnamefont{Esslinger}},
  \bibinfo{journal}{Phys. Rev. Lett.} \textbf{\bibinfo{volume}{87}},
  \bibinfo{pages}{160405} (\bibinfo{year}{2001}).

\bibitem[{\citenamefont{Girardeau}(1960)}]{girardeau}
\bibinfo{author}{\bibfnamefont{M.}~\bibnamefont{Girardeau}},
  \bibinfo{journal}{J. Math. Phys.} \textbf{\bibinfo{volume}{6}},
  \bibinfo{pages}{516} (\bibinfo{year}{1960}).

\bibitem[{\citenamefont{Lieb and Liniger}(1963)}]{lieb}
\bibinfo{author}{\bibfnamefont{E.~H.} \bibnamefont{Lieb}} \bibnamefont{and}
  \bibinfo{author}{\bibfnamefont{W.}~\bibnamefont{Liniger}},
  \bibinfo{journal}{Phys. Rev.} \textbf{\bibinfo{volume}{130}},
  \bibinfo{pages}{1605} (\bibinfo{year}{1963}).

\bibitem[{\citenamefont{Lenard}(1964)}]{lenard}
\bibinfo{author}{\bibfnamefont{A.}~\bibnamefont{Lenard}}, \bibinfo{journal}{J.
  Math. Phys.} \textbf{\bibinfo{volume}{5}}, \bibinfo{pages}{930}
  (\bibinfo{year}{1964}).

\bibitem[{\citenamefont{Lenard}(1966)}]{lenard66}
\bibinfo{author}{\bibfnamefont{A.}~\bibnamefont{Lenard}}, \bibinfo{journal}{J.
  Math. Phys.} \textbf{\bibinfo{volume}{7}}, \bibinfo{pages}{1268}
  (\bibinfo{year}{1966}).

\bibitem[{\citenamefont{Sutherland}(1971{\natexlab{a}})}]{sutherlandpra1}
\bibinfo{author}{\bibfnamefont{B.}~\bibnamefont{Sutherland}},
  \bibinfo{journal}{Phys. Rev. A.} \textbf{\bibinfo{volume}{4}},
  \bibinfo{pages}{2019} (\bibinfo{year}{1971}{\natexlab{a}}).

\bibitem[{\citenamefont{Sutherland}(1971{\natexlab{b}})}]{sutherlandjmp}
\bibinfo{author}{\bibfnamefont{B.}~\bibnamefont{Sutherland}},
  \bibinfo{journal}{J. Math. Phys.} \textbf{\bibinfo{volume}{12}},
  \bibinfo{pages}{246} (\bibinfo{year}{1971}{\natexlab{b}}).

\bibitem[{\citenamefont{Sutherland}(1972)}]{sutherlandpra2}
\bibinfo{author}{\bibfnamefont{B.}~\bibnamefont{Sutherland}},
  \bibinfo{journal}{Phys. Rev. A.} \textbf{\bibinfo{volume}{5}},
  \bibinfo{pages}{1372} (\bibinfo{year}{1972}).


  \bibitem[{\citenamefont{Tracy and Vaidya}(1979{\natexlab{b}})}]{tracyphysrev}
  \bibinfo{author}{\bibfnamefont{H.~G.} \bibnamefont{Vaidya}} \bibnamefont{and}
  \bibinfo{author}{\bibfnamefont{C.~A.} \bibnamefont{Tracy}},
  \bibinfo{journal}{Phys. Rev. Lett.} \textbf{\bibinfo{volume}{42}},
  \bibinfo{pages}{3} (\bibinfo{year}{1979}{\natexlab{b}}).

  \bibitem[{\citenamefont{Tracy and Vaidya}(1979{\natexlab{a}})}]{tracyerrata}
  \bibinfo{author}{\bibfnamefont{H.~G.}\bibnamefont{Vaidya}} \bibnamefont{and}
  \bibinfo{author}{\bibfnamefont{C.~A.} \bibnamefont{Tracy}} ,
  \bibinfo{journal}{Phys. Rev. Lett.} \textbf{\bibinfo{volume}{43}},
  \bibinfo{pages}{1540} (\bibinfo{year}{1979}{\natexlab{a}}).


\bibitem[{\citenamefont{Vaidya and Tracy}(1979)}]{tracy}
\bibinfo{author}{\bibfnamefont{H.~G.} \bibnamefont{Vaidya}} \bibnamefont{and}
  \bibinfo{author}{\bibfnamefont{C.~A.} \bibnamefont{Tracy}},
  \bibinfo{journal}{J. Math. Phys.} \textbf{\bibinfo{volume}{20}},
  \bibinfo{pages}{11} (\bibinfo{year}{1979}).

\bibitem[{\citenamefont{Jimbo et~al.}(1980)\citenamefont{Jimbo, Miwa, Mori, and
  Sato}}]{jimbo}
\bibinfo{author}{\bibfnamefont{M.}~\bibnamefont{Jimbo}},
  \bibinfo{author}{\bibfnamefont{T.}~\bibnamefont{Miwa}},
  \bibinfo{author}{\bibfnamefont{Y.}~\bibnamefont{Mori}}, \bibnamefont{and}
  \bibinfo{author}{\bibfnamefont{M.}~\bibnamefont{Sato}},
  \bibinfo{journal}{Phys. D} \textbf{\bibinfo{volume}{1}}, \bibinfo{pages}{80}
  (\bibinfo{year}{1980}).

\bibitem[{\citenamefont{Dunjko et~al.}(2001)\citenamefont{Dunjko, Lorent, and
  Olshanii}}]{dunjko}
\bibinfo{author}{\bibfnamefont{V.}~\bibnamefont{Dunjko}},
  \bibinfo{author}{\bibfnamefont{V.}~\bibnamefont{Lorent}}, \bibnamefont{and}
  \bibinfo{author}{\bibfnamefont{M.}~\bibnamefont{Olshanii}},
  \bibinfo{journal}{Phys. Rev. Lett.} \textbf{\bibinfo{volume}{86}},
  \bibinfo{pages}{5413} (\bibinfo{year}{2001}).

\bibitem[{\citenamefont{Petrov et~al.}(2000)\citenamefont{Petrov, Shlyapnikov,
  and Walraven}}]{petrov}
\bibinfo{author}{\bibfnamefont{D.~S.} \bibnamefont{Petrov}},
  \bibinfo{author}{\bibfnamefont{G.~V.} \bibnamefont{Shlyapnikov}},
  \bibnamefont{and} \bibinfo{author}{\bibfnamefont{J.~T.~M.}
  \bibnamefont{Walraven}}, \bibinfo{journal}{Phys. Rev. Lett.}
  \textbf{\bibinfo{volume}{85}}, \bibinfo{pages}{3745} (\bibinfo{year}{2000}).

\bibitem[{\citenamefont{Grenander and Szeg\"o}(1958)}]{szegotoeplitz}
\bibinfo{author}{\bibfnamefont{U.}~\bibnamefont{Grenander}} \bibnamefont{and}
  \bibinfo{author}{\bibfnamefont{G.}~\bibnamefont{Szeg\"o}},
  \emph{\bibinfo{title}{Toeplitz forms and their applications}}
  (\bibinfo{publisher}{University of California Press},
  \bibinfo{address}{Berkeley}, \bibinfo{year}{1958}).

\bibitem[{\citenamefont{Forrester et~al.}()\citenamefont{Forrester, Frankel,
  Garoni, and Witte}}]{cmp}
\bibinfo{author}{\bibfnamefont{P.~J.} \bibnamefont{Forrester}},
  \bibinfo{author}{\bibfnamefont{N.~E.} \bibnamefont{Frankel}},
  \bibinfo{author}{\bibfnamefont{T.~M.} \bibnamefont{Garoni}},
  \bibnamefont{and} \bibinfo{author}{\bibfnamefont{N.~S.} \bibnamefont{Witte}},
  \bibinfo{note}{submitted to Commun. Math. Phys.}, \eprint{math-ph/0207005}.

\bibitem[{\citenamefont{Forrester and Witte}()}]{peterandnick}
\bibinfo{author}{\bibfnamefont{P.~J.} \bibnamefont{Forrester}}
  \bibnamefont{and} \bibinfo{author}{\bibfnamefont{N.~S.} \bibnamefont{Witte}},
  \eprint{math-ph/0204008}.

\bibitem[{\citenamefont{Lenard}(1972)}]{lenardpacific}
\bibinfo{author}{\bibfnamefont{A.}~\bibnamefont{Lenard}},
  \bibinfo{journal}{Pacific J. Math.} \textbf{\bibinfo{volume}{42}},
  \bibinfo{pages}{137} (\bibinfo{year}{1972}).

\bibitem[{\citenamefont{Widom}(1973)}]{widom}
\bibinfo{author}{\bibfnamefont{H.}~\bibnamefont{Widom}}, \bibinfo{journal}{Am.
  J. Math.} \textbf{\bibinfo{volume}{95}}, \bibinfo{pages}{333}
  (\bibinfo{year}{1973}).

\bibitem[{\citenamefont{Forrester}()}]{petersbook}
\bibinfo{author}{\bibfnamefont{P.~J.} \bibnamefont{Forrester}},
  \emph{\bibinfo{title}{Log gases and random matrices}},
  \urlprefix\url{http://www.ms.unimelb.edu.au/~matpjf/matpjf.html}.

\bibitem[{\citenamefont{Barnes}(1900)}]{barnes}
\bibinfo{author}{\bibfnamefont{E.~W.} \bibnamefont{Barnes}},
  \bibinfo{journal}{Q. J. Pure Appl. Math.} \textbf{\bibinfo{volume}{31}},
  \bibinfo{pages}{264} (\bibinfo{year}{1900}).

\bibitem[{\citenamefont{Basor and Morrison}(1994)}]{basor}
\bibinfo{author}{\bibfnamefont{E.~L.} \bibnamefont{Basor}} \bibnamefont{and}
  \bibinfo{author}{\bibfnamefont{K.~E.} \bibnamefont{Morrison}},
  \bibinfo{journal}{Linear Algebra Appl.} \textbf{\bibinfo{volume}{202}},
  \bibinfo{pages}{129} (\bibinfo{year}{1994}).

\bibitem[{\citenamefont{Ayoub}(1981)}]{ayoub}
\bibinfo{author}{\bibfnamefont{R.~G.} \bibnamefont{Ayoub}},
  \bibinfo{journal}{Amer. Math. Monthly} \textbf{\bibinfo{volume}{74}},
  \bibinfo{pages}{1067} (\bibinfo{year}{1981}).

\bibitem[{\citenamefont{Dunham}(1999)}]{euler}
\bibinfo{author}{\bibfnamefont{W.}~\bibnamefont{Dunham}},
  \emph{\bibinfo{title}{Euler. The Master of Us All}} (\bibinfo{publisher}{The
  Mathematical Association of America}, \bibinfo{address}{Washington DC},
  \bibinfo{year}{1999}).

\bibitem[{\citenamefont{Prudnikov et~al.}(1992)\citenamefont{Prudnikov,
  Brychkov, and Marichev}}]{prud}
\bibinfo{author}{\bibfnamefont{A.~P.} \bibnamefont{Prudnikov}},
  \bibinfo{author}{\bibfnamefont{Y.~A.} \bibnamefont{Brychkov}},
  \bibnamefont{and} \bibinfo{author}{\bibfnamefont{O.~I.}
  \bibnamefont{Marichev}}, \emph{\bibinfo{title}{Integrals and Series}}
  (\bibinfo{publisher}{Gordon and Breach Science Publishers},
  \bibinfo{address}{New York}, \bibinfo{year}{1992}).

\bibitem[{\citenamefont{Wong}(1989)}]{wong}
\bibinfo{author}{\bibfnamefont{R.}~\bibnamefont{Wong}},
  \emph{\bibinfo{title}{Asymptotic Approximations of Integrals}}
  (\bibinfo{publisher}{Academic Press}, \bibinfo{address}{New York},
  \bibinfo{year}{1989}).

\bibitem[{\citenamefont{Baker and Forrester}(1997)}]{baker}
\bibinfo{author}{\bibfnamefont{T.~H.} \bibnamefont{Baker}} \bibnamefont{and}
  \bibinfo{author}{\bibfnamefont{P.~J.} \bibnamefont{Forrester}},
  \bibinfo{journal}{Commun. Math. Phys.} \textbf{\bibinfo{volume}{188}},
  \bibinfo{pages}{175} (\bibinfo{year}{1997}).

\bibitem[{\citenamefont{Johansson}(1998)}]{johanssonfluctuations}
\bibinfo{author}{\bibfnamefont{K.}~\bibnamefont{Johansson}},
  \bibinfo{journal}{Duke. Math. Journal} \textbf{\bibinfo{volume}{91}},
  \bibinfo{pages}{151} (\bibinfo{year}{1998}).

\bibitem[{\citenamefont{Kalisch and Braak}()}]{kalisch}
\bibinfo{author}{\bibfnamefont{F.}~\bibnamefont{Kalisch}} \bibnamefont{and}
  \bibinfo{author}{\bibfnamefont{D.}~\bibnamefont{Braak}},
  \eprint{cond-mat/0201585}.

\bibitem[{\citenamefont{Lapeyre et~al.}(2002)\citenamefont{Lapeyre, Girardeau,
  and Wright}}]{girardeau8}
\bibinfo{author}{\bibfnamefont{G.~J.} \bibnamefont{Lapeyre}},
  \bibinfo{author}{\bibfnamefont{M.~D.} \bibnamefont{Girardeau}},
  \bibnamefont{and} \bibinfo{author}{\bibfnamefont{E.~M.}
  \bibnamefont{Wright}}, \bibinfo{journal}{Phys. Rev. A.}
  \textbf{\bibinfo{volume}{66}}, \bibinfo{pages}{023606}
  (\bibinfo{year}{2002}).

\bibitem[{\citenamefont{Papenbrock}()}]{papenbrock}
\bibinfo{author}{\bibfnamefont{T.}~\bibnamefont{Papenbrock}},
  \eprint{cond-mat/0209300}.

\bibitem[{\citenamefont{Girardeau et~al.}(2001)\citenamefont{Girardeau, Wright,
  and Triscari}}]{girardeau10}
\bibinfo{author}{\bibfnamefont{M.~D.} \bibnamefont{Girardeau}},
  \bibinfo{author}{\bibfnamefont{E.~M.} \bibnamefont{Wright}},
  \bibnamefont{and} \bibinfo{author}{\bibfnamefont{J.~M.}
  \bibnamefont{Triscari}}, \bibinfo{journal}{Phys. Rev. A.}
  \textbf{\bibinfo{volume}{63}}, \bibinfo{pages}{033601}
  (\bibinfo{year}{2001}).

\bibitem[{\citenamefont{Delves and Mohamed}(1985)}]{delves}
\bibinfo{author}{\bibfnamefont{L.~M.} \bibnamefont{Delves}} \bibnamefont{and}
  \bibinfo{author}{\bibfnamefont{J.~L.} \bibnamefont{Mohamed}},
  \emph{\bibinfo{title}{Computational methods for integral equations}}
  (\bibinfo{publisher}{Cambridge University Press}, \bibinfo{address}{New
  York}, \bibinfo{year}{1985}).

\bibitem[{\citenamefont{Johansson}(1997)}]{johanssoncompactclassicalgroups}
\bibinfo{author}{\bibfnamefont{K.}~\bibnamefont{Johansson}},
  \bibinfo{journal}{Ann. of Math. (2)} \textbf{\bibinfo{volume}{145}},
  \bibinfo{pages}{519} (\bibinfo{year}{1997}).

\bibitem[{\citenamefont{Forrester}(1992)}]{peter2}
\bibinfo{author}{\bibfnamefont{P.~J.} \bibnamefont{Forrester}},
  \bibinfo{journal}{Phys. Lett. A} \textbf{\bibinfo{volume}{163}},
  \bibinfo{pages}{121} (\bibinfo{year}{1992}).

\bibitem[{\citenamefont{Br\'ezin and Hikami}(2000)}]{brezin}
\bibinfo{author}{\bibfnamefont{E.}~\bibnamefont{Br\'ezin}} \bibnamefont{and}
  \bibinfo{author}{\bibfnamefont{S.}~\bibnamefont{Hikami}},
  \bibinfo{journal}{Commun. Math. Phys.} \textbf{\bibinfo{volume}{214}},
  \bibinfo{pages}{111} (\bibinfo{year}{2000}).

\bibitem[{\citenamefont{Strahov and Fyodorev}()}]{strahov}
\bibinfo{author}{\bibfnamefont{E.}~\bibnamefont{Strahov}} \bibnamefont{and}
  \bibinfo{author}{\bibfnamefont{Y.~V.} \bibnamefont{Fyodorev}},
  \eprint{math-ph/0210010}.

\bibitem[{\citenamefont{Stieltjes}(1993)}]{stieltjes}
\bibinfo{author}{\bibfnamefont{T.~J.} \bibnamefont{Stieltjes}},
  \emph{\bibinfo{title}{Oeuvres Compl\'etes}}
  (\bibinfo{publisher}{Springer-Verlag}, \bibinfo{address}{Berlin},
  \bibinfo{year}{1993}).

\bibitem[{\citenamefont{Hardy et~al.}(2000)\citenamefont{Hardy, \surname{Seshu
  Aiyar}, and Wilson}}]{ramanujan}
\bibinfo{editor}{\bibfnamefont{G.~H.} \bibnamefont{Hardy}},
  \bibinfo{editor}{\bibfnamefont{P.~V.} \bibnamefont{\surname{Seshu Aiyar}}},
  \bibnamefont{and} \bibinfo{editor}{\bibfnamefont{B.}~\bibnamefont{Wilson}},
  eds., \emph{\bibinfo{title}{Collected Papers of Srinivasa Ramanujan}}
  (\bibinfo{publisher}{AMS Chelsea Publishing}, \bibinfo{address}{U.S.A.},
  \bibinfo{year}{2000}).

\end{thebibliography}

\end{document}